\documentclass[aps,prd,reprint,superscriptaddress,nofootinbib]{revtex4-1}
\usepackage{amssymb,amsmath,graphicx,hyperref}
\usepackage[utf8]{inputenc}
\usepackage{xcolor} 
\definecolor{green}{rgb}{0,0.69,0.2} 

\def\be{\begin{equation}}
\def\ee{\end{equation}}
\def\beal{\begin{equation}\begin{aligned}}
\def\eeal{\end{aligned}\end{equation}}
\def\bse{\begin{subequations}}
\def\ese{\end{subequations}}
\def\nn{\nonumber}

\def\bra#1{\langle #1|}
\def\ket#1{|#1 \rangle}
\def\braket#1{\langle #1 \rangle}
\def\la{\lambda}

\def\Res_#1{\operatorname*{Res}_{#1}}

\def\tr{\operatorname*{tr}}

\def\ie{i.e. }
\def\eg{e.g. }

\def\eqn#1{Eq.~\eqref{#1}}
\def\eqns#1#2{Eqs.~\eqref{#1} and~\eqref{#2}}

\def\sec#1{Sec.~{\ref{#1}}}

\def\hh{\mathfrak{h}}
\newcommand{\scalegraph}[2]{\vcenter{\hbox{\!\;\includegraphics[scale=#1]{#2.eps}\!\;}}}

\begin{document}

\title{Chiral approach to massive higher spins}

\author{Alexander Ochirov}
\email[]{ochirov@maths.ox.ac.uk}
\affiliation{Mathematical Institute, University of Oxford,
Woodstock Road, Oxford OX2 6GG, U.K.}

\author{Evgeny Skvortsov}
\email[]{evgeny.skvortsov@umons.ac.be}
\affiliation{Service de Physique de l’Univers, Champs et Gravitation,
Universit\'e de Mons, 20 place du Parc, 7000 Mons, Belgium}
\affiliation{Lebedev Institute of Physics,
Leninsky ave. 53, 119991 Moscow, Russia}


\begin{abstract}
We propose a new, chiral description for massive higher-spin particles in four spacetime dimensions, which facilitates the introduction of consistent interactions. As proof of concept, we formulate three theories, in which higher-spin matter is coupled to electrodynamics, non-Abelian gauge theory or gravity. The theories are chiral and have simple Lagrangians, resulting in Feynman rules analogous to those of massive scalars. Starting from these Feynman rules, we derive tree-level scattering amplitudes with two higher-spin matter particles and any number of positive-helicity photons, gluons or gravitons. The amplitudes reproduce the arbitrary-multiplicity results that were obtained via on-shell recursion in a parity-conserving setting, and which chiral and non-chiral theories thus have in common. The presented theories are currently the only examples of consistent interacting field theories with massive higher-spin fields. 
\end{abstract}

\maketitle

\section{Introduction
\label{sec:intro}}

The study of higher-spin fields is a formidable subject.
In the massless case, such fields possess rich gauge symmetry,
starting from the familiar case of electromagnetism,
see \eg \cite{Bekaert:2022poo,Ponomarev:2022vjb} for recent reviews. The standard approach is to introduce
fields of spin~$s$ as symmetric traceless tensors $\Phi^{\mu_1\dots\mu_s}$
or spinors $\Psi^{\mu_1\dots\mu_{s-1/2}}$. 
Massive higher-spin fields \cite{Fierz:1939ix,Singh:1974qz,Singh:1974rc,Zinoviev:2001dt}
are more subtle and require a host of auxiliary fields
that prevent propagation of unphysical degrees of freedom
in interacting theories~\cite{Johnson:1960vt,Velo:1969bt}.

Massive higher-spin particles do exist in nature
as composite states of elementary particles
(see \eg \cite{ParticleDataGroup:2022pth}).
One should therefore be able to describe their physics
by suitable effective field theories
involving an infinite hierarchy of higher-dimensional operators.
Although only a finite subset of such operators contribute
at any given order in the energy-scale cutoff,
it is a laborious task to even list them
in a manner consistent with no ghost propagation.

For these reasons, the space of massive higher-spin theories
seems vast and ridden with obstacles.
It seems plausible, however, that there are hidden gems
among such theories,
as indicated by recent on-shell studies.
For instance, at the level of 3-point scattering amplitudes,
various possible on-shell spinor structures
for an electromagnetically interacting higher-spin particle of mass~$m$
have been classified by Arkani-Hamed, Huang and Huang (AHH),
and one out of them was singled out~\cite{Arkani-Hamed:2017jhn}
due to its relatively tame behavior in the massless limit, namely
\be
{\cal A}(1_{\{a\}},2^{\{b\}}\!,3^+)
=\frac{\!\braket{1_{(a_1} 2^{(b_1}}\cdots\braket{1_{a_{2s})} 2^{b_{2s})}}\!}
      {m^{2s}} {\cal A}(1,2,3^+) .
\label{AHH}
\ee
Here ${\cal A}(1,2,3^+)$ is the positive-helicity photon emission amplitude in scalar quantum electrodynamics (QED),
whereas all higher-spin information is encoded
into two sets of ${\rm SU}(2)$ little-group indices,
$\{a_1,\dots,a_{2s}\}$ and $\{b_1,\dots,b_{2s}\}$,
via $2s$~copies of the same chiral product
$\braket{1_a 2^b} := \epsilon_{\alpha\beta} \bra{1_a}^\alpha \bra{2^b}^\beta$
of massive-momentum spinors.

Interestingly, the AHH amplitude may be extended
\cite{Aoude:2020onz,Lazopoulos:2021mna}
to include $(n-2)$ positive-helicity photons, gluons or gravitons
instead of one,
while still having the same massive-spin structure as \eqn{AHH},
see \eg \eqn{ASSggnPlus} below for gauge theory.
These amplitudes can be derived from mere knowledge
of their factorization limits
by na\"ive use of on-shell (BCFW) recursion~\cite{Britto:2004ap,Britto:2005fq}
--- in exactly the same way that is known to work
for their spin-1/2 counterparts in quantum chromodynamics
(QCD)~\cite{Ochirov:2018uyq}.
Unlike mixed-helicity configurations,
where the same approach produces answers afflicted by unphysical poles
\cite{Arkani-Hamed:2017jhn,Johansson:2019dnu},
the like-helicity amplitudes involving a pair of higher-spin particles
seem absolutely healthy.
Another argument in favor of the like-helicity results is that
they can be derived~\cite{Lazopoulos:2021mna}
using two distinct BCFW constructions:
either complex-shifting two massless momenta~\cite{Britto:2005fq}
or one massless and one massive~\cite{Badger:2005zh,Aoude:2019tzn}.

It is well-known that on-shell recursion fails
when the desired amplitudes have bad boundary behavior, \ie they
do not vanish as the complex BCFW-shift parameter~$z$ is taken to infinity.
This is something that should absolutely be expected
from generic effective theories, where the action is built out of
higher-dimensional vertex operators,
typically with a growing number of derivatives.
However, the existence of healthy $n$-point amplitudes
that are constructible from their factorization limits
(albeit in a restricted helicity sector)
suggests that they should belong to a well-defined massive higher-spin theory
--- one that is intimately related to the minimally coupled scalar theory
and thereby satisfies the boundary-behavior condition
allowing for on-shell recursion.

This is precisely the kind of theories that we present in this paper.
We start with \sec{sec:free}, where we review the common difficulties in higher-spin field theory and show how they are avoided by a new chiral Lagrangian description of a free massive higher-spin field in four dimensions.
Minimal gauge interactions are then added to this theory in \sec{sec:gauge}.
After that, we present the corresponding chiral theory of
gravitationally interacting higher-spin field~\sec{sec:gravity}.
Finally, we conclude by discussing some implications
of our theories in \sec{sec:outro}.

\section{Free higher-spin theory
\label{sec:free}}

In this section we discuss the basics of describing
a free higher-spin particle.
For simplicity, let us temporarily concentrate on integer-spin particles.\footnote{The extension of the discussion of the first few paragraphs
of \sec{sec:free} to half-integer spins is straightforward
and can be found \eg in \cite{Johansson:2019dnu}.
}
In the standard approach~\cite{Singh:1974qz},
integer-spin fields constitute symmetric traceless rank-$s$ tensors $\Phi_{\mu_1\dots\mu_s}$.
For instance, in electromagnetism,
\ie the theory of a free massless spin-1 boson,
the off-shell field is $A_\mu$,
and its connection to the on-shell particle description relies
on the notion of polarization vector $\varepsilon_{p,\mu}^\pm$,
which converts the Lorentz index into the helicity label.
Helicity governs the particle's irreducible representation
of the little group ${\rm U}(1)$ for a given massless momentum~$p_\mu$
and is additive in the sense that the polarization tensor
of \eg massless spin-2 boson (graviton) is simply
$\varepsilon_{p,\mu\nu}^\pm:=\varepsilon_{p,\mu}^\pm \varepsilon_{p,\nu}^\pm$.

For $p^2 = m^2 \neq 0$, one can employ symmetric rank-$2s$ tensors
as irreducible representations of the massive little group ${\rm SU}(2)$
with exactly $(2s+1)$ degrees of freedom.
For $s=1$, we therefore use polarization vector
$\varepsilon_{p,\mu}^{a_1 a_2}=\varepsilon_{p,\mu}^{a_2 a_1}$,
where $a_i$ are fundamental ${\rm SU}(2)$ indices,
whereas free higher-spin fields are naturally expanded
in terms of the polarization tensors
\be
\varepsilon_{p,\mu_1 \ldots \mu_s}^{a_1 \ldots a_{2s}}
:= \varepsilon_{p,\mu_1}^{(a_1 a_2}\cdots
   \varepsilon_{p,\mu_s}^{a_{2s-1} a_{2s})} .
\label{PolTensors}
\ee
Symmetry in the Lorentz indices is then obvious,
and tracelessness $\eta^{\mu_i\mu_j}\varepsilon_{p,\mu_1 \ldots \mu_s} = 0$
follows from the spin-1 orthonormality
\be\!\!
\varepsilon_{p,ab}\cdot\varepsilon_p^{cd}
= -\delta_{(a}^{(c} \delta_{b)}^{d)} ,~~\quad
\varepsilon_{p,\mu ab} := \epsilon_{ac} \epsilon_{bd} \varepsilon_{p,\mu}^{cd}
= (\varepsilon_{p,\mu}^{ab})^* ,\!
\label{PolVectorProperties}
\ee
where have used the fact that ${\rm SU}(2)$ indices are lowered and raised
with the two-dimensional Levi-Civita tensor.

The crucial element of this approach is transversality
$p\cdot\varepsilon_p^{\{a\}}=0$, which is required, roughly speaking,
to adjust the number of degrees of freedom on the Lorentz-index side
to that on the little-group side (3 in the spin-1 case).
In other words, the transversality is required
to ensure the irreducibility of the traceless symmetric tensor
representation of the Lorentz group under Wigner's little group
$\{L \in {\rm SO}(1,3): L^\mu{}_\nu p^\nu=p^\mu\}$.\footnote{More concretely, for a generic Lorentz transformation
$L \in {\rm SO}(1,3)$ we have
\be
L_\mu{}^\nu \varepsilon_{p,\nu}^{ab}
= U^a{}_c(L) U^b{}_d(L) \varepsilon_{(Lp),\mu}^{cd} ,
\label{PolVectorLG}
\ee
where $U(L) \in {\rm SU}(2)$ in principle
depends on the choice of the spin quantization axis for any given $p$.
}

Therefore, the free field equations that one needs to impose on
$(s,s)$-representation tensors are
\be
(\partial^2 + m^2) \Phi_{\mu_1\dots\mu_s} = 0 , \qquad
\partial^{\mu} \Phi_{\mu\mu_2\dots\mu_s} = 0 .
\label{FreeFETensor}
\ee
We have referred to the symmetric traceless tensor representation
by the numbers of its chiral and antichiral ${\rm SL}(2,\mathbb{C})$ indices
upon the application of the spinor map
\be
\Phi_{\alpha_1\dots\alpha_s\dot{\beta}_1\dots\dot{\beta}_s} :=
\Phi_{\mu_1\dots\mu_s} \sigma_{\alpha_1\dot{\beta}_1}^{\mu_1}
\cdots \sigma_{\alpha_s\dot{\beta}_s}^{\mu_s} ,
\label{SpinorMap}
\ee
where $\sigma^\mu = (1,\sigma^1,\sigma^2,\sigma^3)$ are the Pauli matrices.
In fact, it is easy to see in this spinor language that,
although $(s,s)$ is irreducible under
${\rm SL}(2,\mathbb{C}) \cong {\rm SO}(1,3)$,
it is highly reducible under ${\rm SU}(2) \subset {\rm SL}(2,\mathbb{C})$
and decomposes into symmetric ${\rm SU}(2)$ tensors of rank $0,2,\ldots,2s$.

The Klein-Fock-Gordon equation in \eqn{FreeFETensor} can be obtained from a simple Lagrangian.
However, there is no action principle for $s>1$ that also generates the second equation in \eqn{FreeFETensor},
which is required to ensure irreducibility under Wigner's little group, unless a host of auxiliary fields is introduced \cite{Fierz:1939ix}. 
For instance, the Singh-Hagen approach~\cite{Singh:1974qz,Singh:1974rc}
relies on introducing $(s-1)$ symmetric traceless tensor fields
of rank $0,1,\ldots,s-2$.
Alternatively, Zinoviev's approach~\cite{Zinoviev:2001dt}
involves $s$~such fields of rank $0,1,\ldots,s-1$,
that are also subject to the double-trace condition
$\Phi^{\la\mu}{}_{\la\mu\mu_5\dots\mu_{s-k}} = 0$.
All these fields vanish on shell
but serve to ensure the free-field expansion of $\Phi_{\mu_1\dots\mu_s}$
in terms of the physical polarization tensors \eqref{PolTensors}.

Here, our radically simple proposal,
as inspired by the higher-spin amplitudes~\eqref{AHH}
and by chiral higher-spin gravity~\cite{Metsaev:1991mt,Metsaev:1991nb,Ponomarev:2016lrm,Skvortsov:2018jea,Krasnov:2021nsq,Skvortsov:2020wtf,Skvortsov:2020gpn,Sharapov:2022faa,Sharapov:2022awp},
is to \emph{take basic fields $\Phi_{\alpha_1\dots\alpha_{2s}}$
in the chiral $(2s,0)$ representation of the Lorentz group}
instead of $(s,s)$.
As we will shortly see, this essentially trivializes the transition between
the off-shell symmetry group ${\rm SL}(2,\mathbb{C})$
and the on-shell little-group ${\rm SU}(2)$.

To be more specific, we employ the massive spinor-helicity
formalism~\cite{Conde:2016vxs,Conde:2016izb,Arkani-Hamed:2017jhn},
which provides perfect building blocks for this construction.
Namely, we use chiral and antichiral on-shell spinors $\ket{p}$ and $[p|$,
such that\footnote{The Weyl spinor indices are raised and lowered with
two-dimensional Levi-Civita tensors
$\epsilon^{\alpha\beta} = \epsilon^{\dot{\alpha}\dot{\beta}} = \epsilon^{ab}
= \big(\begin{smallmatrix} 0 & 1 \\ -1 & 0 \end{smallmatrix}\big)
= -\epsilon_{\alpha\beta} = -\epsilon_{\dot{\alpha}\dot{\beta}} = -\epsilon_{ab}$,
just as ${\rm SU}(2)$ indices.
Spinor products like
$[12]:=\epsilon^{\dot{\alpha}\dot{\beta}}[1|_{\dot{\alpha}}[2|_{\dot{\beta}}$
or $\braket{1^a 2^b}$, which appeared in \eqn{AHH},
are hence antisymmetric.
The spinors obey
an array of properties~\cite{Arkani-Hamed:2017jhn,Ochirov:2018uyq},
\eg the Weyl/Dirac equation
$p_\mu \sigma^\mu_{\alpha\dot{\beta}} |p]^{\dot{\beta}} = m \ket{p}_\alpha$.
}
\begin{subequations} \begin{align}
\label{SpinorHelicity0}
m=0 \quad \Rightarrow\;\!\!\qquad
\ket{p}_\alpha [p|_{\dot{\beta}} & :=
   p_\mu \sigma_{\alpha\dot{\beta}}^\mu , \\*
\label{SpinorHelicity1}
m \neq 0 \quad \Rightarrow \quad 
\ket{p^a}_\alpha [p_a|_{\dot{\beta}} & :=
   p_\mu \sigma_{\alpha\dot{\beta}}^\mu .
\end{align} \label{SpinorHelicity}%
\end{subequations}
All external wavefunctions of quantum fields
can be built out of these spinors.
For instance, the massive polarization vector
can be constructed as~\cite{Guevara:2018wpp,Chung:2018kqs}
\be
   \varepsilon_{p,\:\!\mu}^{ab}
    = \frac{i \bra{p^{(a}}\sigma_\mu|p^{b)}]}{\sqrt{2}m}
\label{PolVector}
\ee
and all of the desired properties, such as \eqn{PolVectorProperties}
and transversality, follow automatically.\footnote{In particular, the relationship~\eqref{PolVectorLG}
between Lorentz and little-group transformations
for $\varepsilon_{p,\:\!\mu}^{ab}$ is now a simple consequence
of the analogous relations for spinors:
\begin{subequations} \begin{align}
S_\alpha{}^\beta \ket{p^a}_\beta & = U^a{}_b(S) \ket{(Lp)^b}_\alpha , \\*
[p^a|_{\dot{\beta}} (S^{\dagger})^{\dot{\beta}}{}_{\dot{\alpha}} &
= U^a{}_b(S) [(Lp)^b|_{\dot{\alpha}} .
\end{align} \label{MassiveSpinorLG}%
\end{subequations}
Here the little-group transformation~$U$ depends on $S$ instead of $L$,
but they are of course subject to the two-to-one correspondence
${\rm SL}(2,\mathbb{C}) \cong {\rm SO}(1,3)$:
\be
L^\mu{}_\nu = \frac{1}{2} \tr(\bar{\sigma}^\mu S \sigma_\nu S^\dagger) ,
\ee
where $\bar{\sigma}^\mu = (1,-\sigma^1,-\sigma^2,-\sigma^3)$.
}

For a higher-spin field in representation $(2s,0)$,
the free field equations reduce to the Klein-Fock-Gordon equation
\be
(\partial^2 + m^2) \Phi_{\alpha_1\dots\alpha_{2s}} = 0 ,
\label{FreeFE}
\ee
required to define the mass shell.
Indeed, the number of degrees of freedom
no longer needs to be artificially reduced,
so the only thing that we need from the corresponding external wavefunctions
is converting the off-shell Weyl-spinor indices
into the on-shell little-group indices.
This is precisely what the massive spinors~\eqref{SpinorHelicity1}
are good for, so the external higher-spin wavefunctions are
\be
\scalegraph{0.86}{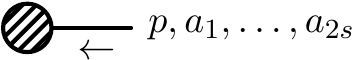} = \frac{1}{m^s}
   \ket{p^{(a_1}}_{\alpha_1}\!\cdots \ket{p^{a_{2s})}}_{\alpha_{2s}} .
\label{FeynRuleExt}
\ee
Here the mass prefactor is needed to absorb the mass dimension,
which is 1/2 for momentum spinors.
The free-field expansion is therefore
\begin{widetext}
\be
\Phi_{\alpha_1\dots\alpha_{2s}}(x)
=\!\int\!\!\frac{\hat{d}^3p}{2p^0}
   \bigg[ 
   \frac{\ket{p^{(a_1}}_{\alpha_1}\!\cdots \ket{p^{a_{2s})}}_{\alpha_{2s}}\!}
        {m^s} a_{a_1\dots a_{2s}}(\vec{\:\!p}) e^{-i p \cdot x}
 + (-1)^{2s}
  \frac{\ket{p_{(a_1}}_{\alpha_1}\!\cdots \ket{p_{a_{2s})}}_{\alpha_{2s}}\!}
        {m^s} a^{\dagger a_1 \dots a_{2s}}(\vec{\:\!p}) e^{i p \cdot x} 
   \bigg] \bigg|_{p^0=\sqrt{\vec{\:\!p}^2+m^2}} ,
\label{FreeField}
\ee
\end{widetext}
where $\hat{d}^3p = (2\pi)^{-3} d^3p$.
We have fixed the signs above so that the outgoing external wavefunctions
\be\!
\scalegraph{0.86}{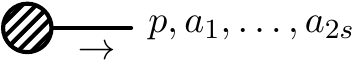} = \frac{(-1)^{2s}\!}{m^{s}}
   \ket{p_{(a_1}}_{\alpha_1}\!\cdots \ket{p_{a_{2s})}}_{\alpha_{2s}}
\label{FeynRuleExtOut}
\ee
are consistent by crossing
with their incoming counterparts~\eqref{FeynRuleExt},
given the momentum-reversal convention
\be
\ket{{-}p} = -\ket{p} , \qquad |{-}p] = |p] .
\label{Crossing}
\ee
In particular, the expansion for the spin-1/2 field $\Phi_\alpha(x)$
coincides with the chiral part of the Majorana field $\Psi_\text{M}(x)$
as written \eg in \cite{Johansson:2019dnu}.
The standard properties
\be
\braket{p^a p^b} = -m \epsilon^{ab} , \qquad
\ket{p^a}_\alpha \bra{p_a}^\beta = -m \delta_\alpha^\beta 
\ee
of the massive spinors are then equivalent to the orthonormality
and completeness relations for spin 1/2.\footnote{For spin $s$, the orthonormality and completeness relations
are explicitly
\begin{subequations} \begin{align}
\label{Orthonormality}
 & \frac{(-1)^{2s}\!}{m^{2s}}
   \braket{p_{(a_1} p^{(b_1}} \cdots \braket{p_{a_{2s)}} p^{b_{2s})}}
    = \delta_{a_1}^{(b_1} \cdots \delta_{a_{2s}}^{b_{2s})} , \\
\label{Completeness} &
\begin{aligned}
   \frac{(-1)^{2s}\!}{m^{2s}}
   \ket{p^{(a_1}}_{\alpha_1}\!\cdots \ket{p^{a_{2s})}}_{\alpha_{2s}}
   \bra{p_{(a_1}}^{\beta_1}\!\cdots \bra{p_{a_{2s})}}^{\beta_{2s}} & \\*
   \qquad \qquad \qquad \qquad \qquad \qquad
    = \delta_{\alpha_1}^{(\beta_1}\!\cdots
      \delta_{\alpha_{2s}}^{\beta_{2s})} .\:&
\end{aligned}
\end{align} 
\end{subequations}
Note that the propagator numerator in \eqn{FeynRuleProp} is consistent with
the completeness relation above.
}

The Lagrangian implying the free field equation~\eqref{FreeFE} is
\be
{\cal L}_0
= \frac{1}{2} (\partial_\mu \Phi^{\alpha_1\dots\alpha_{2s}})
  (\partial^\mu \Phi_{\alpha_1\dots\alpha_{2s}})
- \frac{m^2\!}{2} \Phi^{\alpha_1\dots\alpha_{2s}}
  \Phi_{\alpha_1\dots\alpha_{2s}} .
\label{ActionFree}
\ee
where we treat $\Phi$ as a real/hermitian field.
The reality has a literal meaning in Euclidean or split signature
and should be understood in the sense of \eqn{FreeField}
in Minkowski spacetime.
The corresponding propagator is
\be
\scalegraph{0.86}{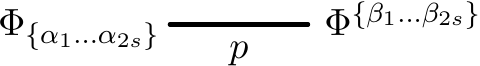}
= \frac{i \delta_{\alpha_1}^{(\beta_1}\!\cdots
          \delta_{\alpha_{2s}}^{\beta_{2s})}}{p^2 - m^2} .
\label{FeynRuleProp}
\ee
Note that, when all indices are raised, the propagator
is automatically antisymmetric for half-integer spins
and symmetric for integer spins.
Indeed, the spin-statistics theorem is also automatically implemented
at the level of (classical) fields:
\beal
\Phi^{\{\alpha\}}(x) \Phi_{\{\alpha\}}(y) &
= (-1)^{2s} \Phi_{\{\alpha\}}(x) \Phi^{\{\alpha\}}(y) \\* &
= \Phi^{\{\alpha\}}(y) \Phi_{\{\alpha\}}(x) ,
\eeal
so they can be manipulated much like scalars.

\section{Gauge theory
\label{sec:gauge}}

In this section we include minimal gauge interactions
in the chiral higher-spin theory~\eqref{ActionFree}.
Since complex conjugation switches between
the chiral and antichiral representations of the Lorentz group,
we choose ${\rm O}(N)$ as the generic gauge group,
which encompasses other interesting cases,
such as ${\rm SU}(N) \subset {\rm O}(2N)$.

We take the Lagrangian to be simply
\be
{\cal L}_g
= \frac{1}{2} (D_\mu \Phi^{\{\alpha\}})_i (D^\mu \Phi_{\{\alpha\}})_i
- \frac{m^2\!}{2} \Phi_i^{\{\alpha\}} \Phi_{i\{\alpha\}} ,
\label{ActionGauge}
\ee
where the covariant derivative
\be\!\!\!
D_\mu := \partial_\mu + g A_\mu ,\!\quad
A_\mu = A_\mu^A t^A ,\!\quad [t^A,t^B] = f^{ABC} t^C\!.
\label{CovDerivative}
\ee
involves antisymmetric ${\rm O}(N)$ generators
in the antihermitian convention $t^A_{ij} = -t^A_{ji}$.
The higher-spin field interacts with the gauge field
via the 3-point vertex
\begin{subequations}
\be
\scalegraph{0.86}{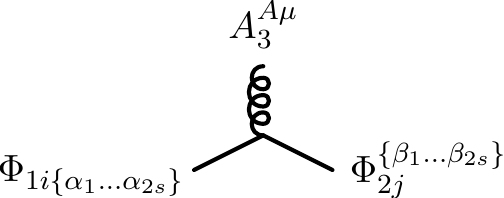}\!\!\!\!\!= g t^A_{ij}
\delta_{\alpha_1}^{(\beta_1}\!\cdots \delta_{\alpha_{2s}}^{\beta_{2s})}
(p_1-p_2)^\mu ,
\label{FeynRuleVSSg}
\ee
and the 4-point vertex
\be\!\!
\scalegraph{0.86}{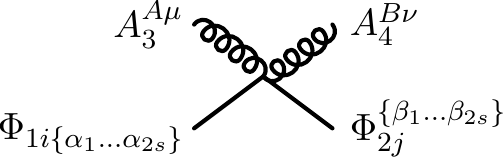}\!\!\!\!\!\!\
\begin{aligned}
 = -ig^2 [t^A_{ik} t^B_{kj} + t^B_{ik} t^A_{kj}] & \\ \times
   \delta_{\alpha_1}^{(\beta_1}\!\cdots \delta_{\alpha_{2s}}^{\beta_{2s})}
   \eta^{\mu\nu} & .\!\!
\end{aligned}
\label{FeynRuleVSSgg}
\ee
\label{FeynRulesQCD}%
\end{subequations}
The crucial feature of these vertices is that
they equal those of a scalar minimally coupled to a gauge field
times the $(2s,0)$-identity operator.
This guarantees that all of the tree-level scattering amplitudes
in this theory will essentially factorize onto those of massive scalar QCD.
For a single pair of higher-spin particles, we have
\be
{\cal A}(1_{\{a\}},2^{\{b\}}\!,3^{h_3}\!,\ldots,n^{h_n}\!) 
 = \frac{\braket{1_a 2^b}^{\odot 2s}\!}{m^{2s}}
   {\cal A}(1,2,3^{h_3}\!,\ldots,n^{h_n}\!) .
\label{HigherSpinFactor}
\ee
Here
${\cal A}(1,2,3^{h_3}\!,\ldots,n^{h_n}\!)$
is the $(n-2)$-photon emission amplitude 
in scalar QED,
and $\odot$ denotes the symmetrized tensor product for massive spinors:
\be
\braket{1_a 2^b}^{\odot 2s} :=
   \braket{1_{(a_1} 2^{(b_1}} \braket{1_{a_2} 2^{b_2}} \cdots
   \braket{1_{a_{2s})} 2^{b_{2s})}} .
\ee
Since there are no vertices with more than two massive fields,
we can easily write similar relations for amplitudes with
multiple pairs of higher-spin particles. For instance,
\begin{align}
 & {\cal A}(1_{\{a\}},2^{\{b\}}\!,3_{\{c\}}\!,4^{\{d\}}\!,
            5^{h_5}\!,\ldots,n^{h_n}\!) \nn \\ &
\label{HigherSpinFactors}
 = \frac{\braket{1_a 2^b}^{\odot 2s}
         \braket{3_c 4^d}^{\odot 2s}\!}{m^{4s}}
   {\cal A}(1,2,{\color{green}3},{\color{green}4},
            5^{h_5}\!,\ldots,n^{h_n}\!) \\* &\,\quad
 + (-1)^{2s}
   \frac{\braket{1_a 4^d}^{\odot 2s}
         \braket{3_c 2^b}^{\odot 2s}\!}{m^{4s}}
   {\cal A}(1,4,{\color{green}3},{\color{green}2},
            5^{h_5}\!,\ldots,n^{h_n}\!) , \nn
\end{align}
where the amplitudes on the right-hand side involve
distinctly flavored scalars with one flavor shown in green.

This is clearly a very chiral theory on the matter side.
Indeed, the 3-point amplitudes that follow
from the vertex~\eqref{FeynRuleVSSg} are explicitly\footnote{As usual,
massless polarization vectors
can be constructed from massless momentum spinors~\eqref{SpinorHelicity0} as
\be
\varepsilon_{p,\mu}^+ = \frac{\bra{q}\sigma_\mu|p]}{\sqrt{2} \braket{qp}} ,
\qquad
\varepsilon_{p,\mu}^- = \frac{\bra{p}\sigma_\mu|q]}{\sqrt{2} [pq]} .
\label{PolVectors}
\ee
In contrast to \eqn{PolVector},
the presence of extraneous reference spinors $\bra{q}$ and $|q]$
encodes the linearized gauge freedom
$\varepsilon_{p,\mu} \to \varepsilon_{p,\mu} + f p_\mu$.
}
\beal
{\cal A}(1^{\{a\}}_i\!,2^{\{b\}}_j\!,3^\pm_A) &
 = 2gt^A_{ij} \frac{\braket{1^a 2^b}^{\odot 2s}\!}{m^{2s}}
   (p_1\cdot \varepsilon_3^\pm) .
\label{A3}
\eeal
Comparing with the original parity-conserving
AHH amplitudes~\cite{Arkani-Hamed:2017jhn} for gauge theory,\footnote{In \cite{Johansson:2019dnu,Lazopoulos:2021mna}
spin-$s$ amplitudes, such as~\eqref{A3} and~\eqref{ASSggnPlus},
carried an additional prefactor $(-1)^{\lfloor s \rfloor}$ due to
the massive polarization vector~\eqref{PolVector}
being conventionally spacelike,
which is irrelevant for our chiral theories here.
}
\begin{subequations} \begin{align}
\label{AHH3plus}
{\cal A}_\text{AHH}(1^{\{a\}}_i\!,2^{\{b\}}_j\!,3^+_A) &
 = 2 g t^A_{ij} \frac{\braket{1^a 2^b}^{\odot 2s}\!}{m^{2s}}
   (p_1\cdot \varepsilon_3^+) , \\*
\label{AHH3minus}
{\cal A}_\text{AHH}(1^{\{a\}}_i\!,2^{\{b\}}_j\!,3^-_A) &
 = 2 g t^A_{ij} \frac{[1^a 2^b]^{\odot 2s}\!}{m^{2s}}
  (p_1\cdot \varepsilon_3^-) ,
\end{align} \label{AHH3}%
\end{subequations}
we observe agreement for positive helicity and mismatch for negative helicity.
This reflects the intrinsic chirality of the higher-spin theory~\eqref{ActionGauge},
see \sec{sec:outro} for the discussion of how parity could be restored.
Here we concentrate on the (incoming) positive-helicity gluons.
Such states correspond to
\emph{the self-dual sector of Yang-Mills theory}
in the sense that the part of the linearized field strength
$F_{\alpha\dot{\alpha}, \beta\dot{\beta}}
:= F_{\mu\nu} \sigma^{[\mu}_{\alpha\dot{\alpha}}
   \sigma^{\nu]}_{\beta\dot{\beta}}
 = F_{\alpha\beta}^- \epsilon_{\dot{\alpha}\dot{\beta}}
 + \tilde{F}_{\dot{\alpha}\dot{\beta}}^+ \epsilon_{\alpha\beta}$,
which gets Wick-contracted with them, is
\be
\tilde{F}^+_{\dot{\alpha}\dot{\beta}}(p) = i 2\sqrt{2}\pi
   [p|_{\dot{\alpha}} [p|_{\dot{\beta}}
   \big[ \delta^+(p^2) a_+(\vec{\:\!p})
       - \delta^-(p^2) a_-^\dagger(-\vec{\:\!p}) \big]
\label{FieldStrengthSD}
\ee
and satisfies
$\frac{1}{2} \epsilon_{\mu\nu\rho\sigma} F^{+\rho\sigma} = i F_{\mu\nu}^+$.
Much is known about the self-dual sector~\cite{Ward:1977ta,Parkes:1992rz,Siegel:1992xp,Chalmers:1996rq,Rosly:1996vr,Abou-Zeid:2005zfo,Adamo:2020syc,Adamo:2020yzi}.
Strictly speaking, self-dual Yang-Mills theory (SDYM)
includes both gluonic helicities,
and in the on-shell approach one may deal with it
simply by switching off one of the 3-gluon amplitudes:
\begin{subequations} \begin{align}
\label{YM3ptMHVb}
{\cal A}(1^+_A,2^+_B,3^-_C) & = -\sqrt{2}gf^{ABC}
   \frac{[12]^3}{[23] [31]} , \\
\label{YM3ptMHV}
{\cal A}(1^-_A,2^-_B,3^+_C) & = \sqrt{2}g f^{ABC}
   \frac{\braket{12}^3}{\braket{23} \braket{31}}
~\underset{\text{SDYM}}{\Rightarrow}~ 0 .
\end{align} \label{YM3pt}%
\end{subequations}

In full Yang-Mills theory,
${\cal L}_\text{YM} = -\frac{1}{4} F_{\mu\nu}^A F^{A\mu\nu}$,
both amplitudes~\eqref{YM3pt} appear in factorization limits.
Even if we couple it to our chiral higher-spin matter~\eqref{ActionGauge},
the existence of BCFW shifts with good boundary behavior
is guaranteed \cite{Badger:2005zh}
by the scalar version of this theory
\eg if we choose to shift only gluonic momenta.\footnote{The simplest argument
in favor of good boundary behavior
in massive scalar QCD starts with pure Yang-Mills theory in higher dimensions,
for which such shifts do exist, as discussed \eg in \cite{ArkaniHamed:2008yf}.
After reducing it to four dimensions and leaving only the massless spectrum,
the extra-dimensional gluonic states behave as adjoint-representation scalars,
whereas the low-energy manipulation of introducing masses
in their propagators cannot change their boundary behavior,
which is a high-energy feature.
}
The ensuing on-shell recursion relations \cite{Britto:2004ap,Britto:2005fq}
allow to build up the entire tree-level scattering matrix,
which will by construction be related to that of scalar QCD
via such identities as \eqns{HigherSpinFactor}{HigherSpinFactors}.

In this theory, there is a subset of amplitudes such that
only amplitudes~\eqref{AHH3plus} and~\eqref{YM3ptMHVb}
appear in their factorization limits.
Such amplitudes coincide with those
in SDYM coupled to our massive higher-spin matter~\eqref{ActionGauge}.
In this chiral theory,
all purely gluonic amplitudes automatically vanish at tree level
except for \eqn{YM3ptMHVb}.
Amplitudes with matter do not vanish but
are significantly simpler than those in the full theory
due to the integrability of the self-dual gauge interaction.
The non-trivial overlap of the full and the self-dual theories
consists of the amplitudes involving a single pair of massive higher-spin particles,
to which we have referred in the introduction.
Their color-ordered\footnote{Color ordering used in \eqn{ASSggnPlus}
assumes the standard convention~\cite{Dixon:1996wi}
favoring hermitian generators $T^A := i\sqrt{2} t^A = (T^A)^\dagger$,
which additionally absorb factors of $\sqrt{2}$.
As usual with color ordering, we have relied on the possibility
to put all particles in the adjoint representation of an
$\mathfrak{su}(N)$ (sub)algebra.
}
versions are explicitly~\cite{Lazopoulos:2021mna}
\beal
A(1^{\{a\}}\!,2^+\!,3^+\!,\dots,(n-1)^+\!,n^{\{b\}}) 
 = \frac{i \braket{1^a n^b}^{\odot 2s}\!}{m^{2s-2}} \quad & \\ \times
   \frac{ [2| \prod_{j=2}^{n-3}\!
          \big\{\!\!\not{\!\!P}_{12 \dots j}\!\not{\!p}_{j+1}
               + (s_{12 \dots j}-m^2) \big\} |n\!-\!1] }
        { \prod_{j=2}^{n-2} \braket{j|j\!+\!1} (s_{12 \dots j}-m^2) } & .\!
\label{ASSggnPlus}
\eeal
Here we have written $P_{12 \dots j} = p_1 + \ldots + p_j$ for momentum sums
and $s_{12 \dots j}$ for their Lorentz squares.
The factors involving slashed matrices in the numerator are
$\big\{(P_{12 \dots j}^\mu \bar{\sigma}_\mu^{\dot{\alpha}\gamma})(p_{j+1}^\nu \sigma_{\nu,\gamma\dot{\beta}})
+(s_{12 \dots j}-m^2) \delta^{\dot{\alpha}}_{\dot{\beta}}\big\}$,
and their order of multiplication is such that
$j$ increases from left to right.
The amplitudes~\eqref{ASSggnPlus} are consistent
with those for massive scalars first derived in \cite{Ferrario:2006np},
as well as with those for massive quarks~\cite{Schwinn:2007ee,Ochirov:2018uyq}
and gauged spin-1 matter obtained from Yang-Mills theory
via the Higgs mechanism~\cite{Ballav:2021ahg}.
We have also performed additional numerical checks through 8 points that the Feynman rules~\eqref{FeynRuleVSSg} and~\eqref{FeynRuleVSSgg}, in combination with the standard 3- and 4-gluon vertices, give the same answers as the formula~\eqref{ASSggnPlus}.

In the important special case of ${\rm SO}(2)$,
for which $f^{ABC}=0$ and $t^{A=1}_{ij}=\epsilon^{ij}$,
we combine the fields into
\be
\Phi^{\{\alpha\}}\!:=
\tfrac{1}{\sqrt{2}} \big( \Phi_{j=1}^{\{\alpha\}}
 + i \Phi_{j=2}^{\{\alpha\}} \big) , \quad
\widetilde{\Phi}^{\{\alpha\}}\!:=
\tfrac{1}{\sqrt{2}} \big( \Phi_{j=1}^{\{\alpha\}}
 - i \Phi_{j=2}^{\{\alpha\}} \big) ,
\ee
where the charge conjugation avoids the chirality switch.
The resulting chiral higher-spin QED Lagrangian is
\be
{\cal L}_Q
= \widetilde{(D_\mu \Phi^{\{\alpha\}})} (D^\mu \Phi_{\{\alpha\}})
- m^2 \widetilde{\Phi}^{\{\alpha\}} \Phi_{\{\alpha\}} ,
\label{ActionQED}
\ee
where the covariant derivative
$D_\mu := \partial_\mu - i Q A_\mu$ involves the coupling constant renamed to $Q$.
A straightforward application of the resulting Feynman rules
\begin{align}
\label{FeynRulesQED}
\scalegraph{0.86}{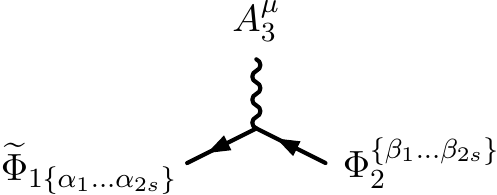}\!\!\!\!\!& = iQ
\delta_{\alpha_1}^{(\beta_1}\!\cdots \delta_{\alpha_{2s}}^{\beta_{2s})}
(p_2-p_1)^\mu , \\
\scalegraph{0.86}{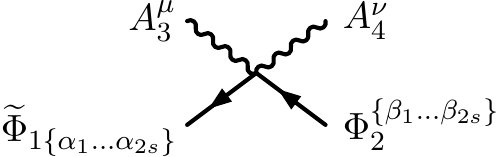}\!\!\!& =
2iQ^2 \delta_{\alpha_1}^{(\beta_1}\!\cdots \delta_{\alpha_{2s}}^{\beta_{2s})} \eta^{\mu\nu} , \nn
\end{align}
leads to the explicit amplitudes (see \eg \cite{Badger:2008rn})
\begin{align}
\label{ASSggnPlusQED}
 & {\cal A}(1^{\{a\}}\!,2^+\!,3^+\!,\dots,(n-1)^+\!,n^{\{b\}})
 = -i(2Q)^{n-2} \\ &
   \times \frac{\braket{1^a n^b}^{\odot 2s}\!}{m^{2s}}\!\!\!
   \sum_{\sigma \in S_{n-2}(\{2,3,\dots,n-1\})}\!
   \frac{ \prod_{j=2}^{n-1} (P_{1\sigma(2)\dots\sigma(j)}
                             \cdot\varepsilon_{\sigma(j)}^+) }
        { \prod_{j=2}^{n-2} (s_{1\sigma(2)\dots\sigma(j)} - m^2) } . \nn
\end{align}
We have checked numerically through 8 points
that they are consistent with the color-dressed analogue of \eqn{ASSggnPlus}
under the usual QCD-to-QED projection
\beal
 & {\cal A}(1^{\{a\}}\!,2,3,\dots,n-1,n^{\{b\}}) \\ &
 = \big(\sqrt{2}Q\big)^{n-2}\!\!\!\!\!
   \sum_{\sigma \in S_{n-2}(\{2,3,\dots,n-1\})}\!\!\!\!\!
   A(1^{\{a\}}\!,\sigma,n^{\{b\}}) .
\label{QCD2QED}
\eeal
Note that all gluonic factorization channels non-trivially cancel here,
leaving only the massive ones as in \eqn{ASSggnPlusQED}.

\section{Gravity
\label{sec:gravity}}

In this section we minimally couple our chiral higher-spin theory
to gravity. The Lagrangian is
\be\!\!
{\cal L}_G = \sqrt{-g} \bigg\{
   \frac{1}{2} (\nabla_\mu \Phi^{\{\alpha\}}) (\nabla^\mu \Phi_{\{\alpha\}})
 - \frac{m^2\!}{2} \Phi^{\{\alpha\}} \Phi_{\{\alpha\}}
   \bigg\} ,\!
\label{ActionGravity}
\ee
where we have included the metric dependency
of the integration measure.
The covariant derivatives may in general act
on the Lorentz indices via the spin connection:
\be
\nabla_\mu \Phi_{\alpha_1\dots\alpha_{2s}}
= \partial_\mu \Phi_{\alpha_1\dots\alpha_{2s}}
+ 2s\;\!\omega_{\mu,(\alpha_1}{}^\beta \Phi_{\alpha_2\dots\alpha_{2s})\beta} .
\label{CovDerivative2}
\ee
This prevents the vertices from factorizing onto those of the massive-scalar theory.
So unlike in the gauge-theoretic case, we cannot make
helicity-independent statements of the type~\eqref{HigherSpinFactor}.

However, much like in gauge theory,
there is a clear connection between positive-helicity gravitons
and the self-duality condition of the Riemann tensor.
Moreover, it is equally true for the spin connection:
the chiral part of the spin connection is known to vanish
in self-dual gravity (SDGR)
\cite{Penrose:1976js,Penrose:1976jq,Capovilla:1991qb,Abou-Zeid:2005zfo}:
\be
\omega_{\mu,\alpha}{}^\beta := \frac{1}{2} \omega_\mu{}^{\hat{\nu}\hat{\rho}}
   \sigma_{\hat{\nu},\alpha\dot{\gamma}}
   \bar{\sigma}_{\hat{\rho}}^{\dot{\gamma}\beta}
   ~\underset{\text{SDGR}}{\Rightarrow}~ 0 ,
\label{SpinConnectionASD}
\ee
where the frame indices are displayed with hats.
So restricting to the self-dual gravitational sector,
we can easily write the 3-point interaction vertex
\be\!\!
\scalegraph{0.86}{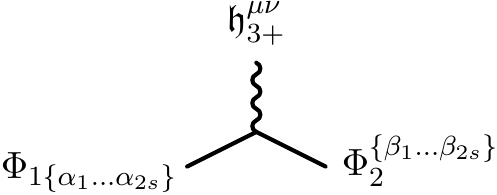}
\begin{aligned} & \\
\!=\,&i\kappa
\delta_{\alpha_1}^{(\beta_1}\!\cdots \delta_{\alpha_{2s}}^{\beta_{2s})} \\
&\!\!\times\!\bigg[ p_2^{(\mu} p_2^{\nu)} + \frac{m^2}{2} \eta^{\mu\nu} \bigg] ,\!\!
\end{aligned}
\label{FeynRuleVSSG}
\ee
\eg in terms of the perturbation of the ``gothic inverse metric''
$\sqrt{-g}\:\!g^{\mu\nu} = \eta^{\mu\nu} - \kappa \mathfrak{h}^{\mu\nu}$
\cite{Landau:1982dva,Poisson:2014}
and the coupling constant $\kappa = \sqrt{32\pi G_\text{(Newton)}}$.
In this formulation of self-dual gravitational perturbation theory,
all vertices involving two massive fields and multiple gravitons
come exclusively from the mass term due to
\be
\label{MetricDetSqrt}
\sqrt{-g} = 1 - \frac{\kappa}{2} \mathfrak{h}
    + \frac{\kappa^2}{8} \big( \hh^2 - 2 \hh^{\mu\nu} \hh_{\mu\nu} \big)
    + O(\kappa^3) ,
\ee
where $\hh:= \hh^{\mu\nu} \eta_{\mu\nu}$.
In any case, in view of the vanishing
anti-self-dual spin connection~\eqref{SpinConnectionASD},
all of the Feynman rules clearly factorize onto those for a massive scalar.

Therefore, for positive-helicity gravitational amplitudes
involving two higher-spin matter particles,
we do have the factorization property
\be
{\cal M}(1_{\{a\}},2^{\{b\}}\!,3^+\!,\ldots,n^+\!) 
 = \frac{\braket{1_a 2^b}^{\odot 2s}\!}{m^{2s}}
   {\cal M}(1,2,3^+\!,\ldots,n^+) .
\label{HigherSpinFactor2}
\ee
Equivalently, these amplitudes are constructible~\cite{Aoude:2020onz}
from their factorization limits via on-shell recursion
starting from the two 3-point amplitudes
\be
{\cal M}(1^{\{a\}}\!,2^{\{b\}}\!,3^{+})
 =-i\kappa \frac{\braket{1^a 2^b}^{\odot 2s}\!}{m^{2s}}
   (p_1\cdot \varepsilon_3^+)^2
\label{AHH3GRplus}
\ee
and
\begin{subequations}
\be
\label{GR3ptMHVb}
{\cal M}(1^+\!,2^+\!,3^-) = -\frac{i\kappa}{2}
   \frac{[12]^6}{[23]^2 [31]^2} ,
\ee
but with no reference to the second 3-graviton amplitude
\be
\label{GR3ptMHV}
{\cal M}(1^-\!,2^-\!,3^+) = -\frac{i\kappa}{2}
   \frac{\braket{12}^6}{\braket{23}^2 \braket{31}^2}   
   ~\underset{\text{SDGR}}{\Rightarrow}~ 0 .
\ee \label{GR3pt}%
\end{subequations}

However, perhaps the easiest way to express
the amplitudes~\eqref{HigherSpinFactor2} is by using
the Kawai-Lewellen-Tye-style double copy~\cite{Kawai:1985xq,Bern:2008qj},
which is known~\cite{Bjerrum-Bohr:2013bxa,Naculich:2014naa,Plefka:2019wyg}
to hold for minimally coupled massive scalars.
The gravitational all-plus amplitudes are thus given
as a bilinear combination of their gauge-theoretic color-ordered counterparts~\eqref{ASSggnPlus}:
\beal
{\cal M}(1^{\{a\}},2^{\{b\}}\!,3^+\!,\ldots,n^+)
=-i\Big(\frac{\kappa}{2}\Big)^{\!n-2} \frac{\braket{1^a 2^b}^{\odot 2s}\!}{m^{2s}} & \\
\times \sum_{\sigma, \tau} 
A(1,2,3^+\!,\sigma) A(2,1,3^+\!,\tau)
\prod_{i=1}^{n-3}
\sum_{j \in X_{\sigma,\tau}^i}
s_{\sigma(i) j} & .
\label{MSSGGnPlus}
\eeal
Here the outer summation is over $\sigma,\tau \in S_{n-3}(\{4,\dots,n\})$, 
$t_i$ is defined as the position of $\sigma(i)$ within permutation $\tau$, \ie
$\tau(t_i) = \sigma(i)$,
and $X_{\sigma,\tau}^i:=\{3,\sigma(1),\dots,\sigma(i-1)\}
\cap \{3,\tau(1),\dots,\tau(t_i-1)\}$. 
This expression is also consistent
with the double copy for massive matter with spin~\cite{Johansson:2015oia,delaCruz:2016wbr,Brown:2018wss,Johansson:2019dnu,Bautista:2019evw}.

\section{Summary and discussion
\label{sec:outro}}

We have presented a chiral approach to massive higher-spin fields
and formulated three theories, in which such fields
interact strongly, electromagnetically or gravitationally.
We have focused on the self-dual sectors of these interactions,
where scattering amplitudes coincide with those derived previously
assuming parity conservation.
On the one hand, all three theories may be consistently truncated down
to their respective self-dual sectors.
On the other hand, they may be extended
by introducing additional vertex operators to their Lagrangians.

The problem of healthy interactions
for higher-spin fields is important for various reasons.
Higher-spin states might be indispensable for building consistent
quantum gravity models, as indicated by string theory,
the AdS/CFT correspondence and higher-spin gravities.
Moreover, massive higher-spin particles can model many
real physical systems within the effective field theory approach.
A suitable implementation of the classical limit
\cite{Guevara:2017csg,Guevara:2018wpp,Chung:2018kqs,Maybee:2019jus,Guevara:2019fsj,Arkani-Hamed:2019ymq,Bern:2020buy,Aoude:2020onz,Aoude:2021oqj}
even allows to model gravitational dynamics of spinning black holes or other compact objects.

For marginal spins, the problem of consistent interactions
has a solution for massive spin-1 fields,
which always result from a spontaneously broken Yang-Mills theory,
and consistent massive spin-2 theories are known as massive gravity~\cite{Bergshoeff:2009hq,deRham:2010kj,Hassan:2011zd,deRham:2014zqa}.
For higher spins, this problem can be roughly subdivided into two:
self-interactions of massive higher-spin fields
and their gauge or gravitational interactions.
The latter is more important for such applications as black-hole scattering.
No solution to either problem has existed beyond spin-2,
and the present paper aims to provide a new way forward.

Let us compare our chiral-field approach to other descriptions
of massive higher-spin fields.
In the Singh-Hagen formulation~\cite{Singh:1974qz,Singh:1974rc},
the auxiliary fields are fine-tuned to give enough differential consequences
of the field equations to guarantee the correct number
of physical degrees of freedom (p.d.o.f.),
which is almost equivalent to a tedious analysis of Hamiltonian constraints.
Zinoviev's approach~\cite{Zinoviev:2001dt}
has a more transparent pattern of auxiliary fields
intertwined by gauge symmetries \`a la St\"uckelberg,
and the remaining challenge is of a purely technical nature \cite{Zinoviev:2006im,Zinoviev:2008ck,Zinoviev:2009hu,Zinoviev:2010av,Zinoviev:2011fv,Buchbinder:2012xa}.
If one is concerned only with field equations,
more economic approaches to control the number of p.d.o.f. can be applied,
see \eg \cite{Buchbinder:2005ua,Buchbinder:2007ix,Kaparulin:2012px,Kazinski:2005eb}.
Furthermore, the light-front approach starts out with the correct p.d.o.f.,
see \eg \cite{Metsaev:2005ar,Metsaev:2007rn}
and especially \cite{Metsaev:2022yvb} for the four-dimensional case,
but the study beyond the cubic order can still be quite laborious. For twistorial constructions involving mass see \eg \cite{Albonico:2022pmd} and the references therein.

Our present proposal for massive higher-spin particles is inspired
by the recent attempts to bootstrap their on-shell amplitudes with
as little off-shell input as possible \cite{Arkani-Hamed:2017jhn,Johansson:2019dnu,Falkowski:2020aso,Chiodaroli:2021eug},
which works surprisingly well for like-helicity configurations
of the force-carrier bosons~\cite{Aoude:2020onz,Lazopoulos:2021mna}.
The chiral description of massive higher-spin interactions
is very close in spirit to the ideas originating from twistor
theory \cite{Penrose:1965am,Hughston:1979tq,Eastwood:1981jy,Woodhouse:1985id,Huggett:1986fs}
and from the covariantization of chiral higher-spin gravity
\cite{Krasnov:2021nsq,Skvortsov:2022syz,Sharapov:2022faa,Sharapov:2022awp}, where a chiral field $\Phi_{\alpha_1\dots\alpha_{2s}}$ is used
to describe massless spin-$s$ states. 
It is also close to the light-cone gauge in not having redundant degrees of freedom, the advantage being in maintaining manifest Lorentz symmetry.
For spin-$1/2$ fields, the chiral description can be understood as
integrating out half of the fermion components out of QCD
\cite{Chalmers:1997ui}, and likewise for spin-1 \cite{Chalmers:2001cy}
after first integrating in an auxiliary field.
However, no such interpretation is available for higher spin fields,
while a parent action~\cite{Fradkin:1984ai} may still exist.

Having presented the three chiral theories,
we can already comment on their extensions.
Similarly to the self-dual theories, our theories,
while being consistent on their own,
generate a subset of the amplitudes of their would-be non-chiral completions.
In the case of SDYM (with matter) and SDGR the completions to full Yang-Mills theory (with matter) and gravity are known
\cite{Chalmers:1996rq,Chalmers:2001cy,Krasnov:2016emc},
and it would be interesting to find such completions for our proposals.
Aiming at restoring parity can also be motivated
by the applications to spinning black holes.
Indeed, the gravitational AHH amplitude~\eqref{AHH3GRplus}
was shown~\cite{Guevara:2018wpp,Chung:2018kqs}
to contain the same spin-induced multipole moments
as a spinning black hole~\cite{Levi:2015msa,Porto:2016pyg,Vines:2017hyw},
but only in combination with its antichiral version
\be
{\cal M}_\text{AHH}(1^{\{a\}}\!,2^{\{b\}}\!,3^-)
=-i\kappa \frac{[1^a 2^b]^{\odot 2s}\!}{m^{2s}} (p_1\cdot \varepsilon_3^-)^2 .
\label{AHH3GRminus}
\ee
On the classical-gravity side,
the tower of multipole moments can be extracted
directly from the linearized Kerr solution
\cite{Chung:2018kqs,Guevara:2020xjx}.
For a spin-$s$ particle described by \eqns{AHH3GRplus}{AHH3GRminus},
it appears truncated down to the $2s$-pole
with correct black-hole multipole coefficients ---
even before the classical limit is taken
\cite{Maybee:2019jus,Guevara:2019fsj,Arkani-Hamed:2019ymq,Bern:2020buy,Aoude:2020onz,Aoude:2021oqj}.
(Interestingly, this ``spin universality''
property~\cite{Holstein:2008sw,Holstein:2008sx,Vaidya:2014kza}
is not obeyed by all three-point amplitudes~\cite{Aoude:2021oqj},
as recently demonstrated~\cite{Cangemi:2022abk}
for the leading Regge states of the open and closed superstring.)

For simplicity, let us consider the gauge-theoretic case,
where the classical analogue of the spinning black hole is
known as the $\sqrt{\text{Kerr}}$ solution~\cite{Arkani-Hamed:2019ymq}.
We can easily write the general form of the interaction terms
that could modify the 3-point amplitude \eqref{AHH3minus}
without spoiling \eqref{AHH3plus}:
\begin{align}
(D_{\alpha_1\dot{\gamma}_1}\!\cdots D_{\alpha_k\dot{\gamma}_k} &
\Phi^{{\alpha_1}\dots\alpha_{2s}})
\epsilon_{\alpha_{k+1}\beta_{k+1}}\!\cdots \epsilon_{\alpha_{2s-1}\beta_{2s-1}}
F_{\alpha_{2s}\beta_{2s}} \nn \\ \times &
(D_{\beta_1}{}^{\dot{\gamma}_1}\!\cdots D_{\beta_k}{}^{\dot{\gamma}_k} 
\Phi^{{\beta_1}\dots\beta_{2s}}) .
\end{align}
A single such term $\Phi^\alpha F_{\alpha\beta} \Phi^\beta$ is in fact known to be sufficient for restoring parity to the entire spin-1/2 theory,
which then constitutes a chiral formulation of QCD with massive quarks
due to Chalmers and Siegel \cite{Chalmers:1997ui}.
As for higher spins, we will explore
their gauge and gravitational interactions in more detail elsewhere.

\begin{acknowledgments}
We thank Rafael Aoude, Lucile Cangemi, Marco Chiodaroli, Henrik Johansson and Paolo Pichini
for stimulating conversations and collaboration on related projects.
We are also grateful to Lionel Mason, Dmitry Ponomarev and Yael Shadmi for useful discussions.
AO's research is funded by the STFC grant ST/T000864/1.
The work of ES is partially supported by the European Research Council (ERC)
under the European Union’s Horizon 2020 research and innovation programme
(grant agreement No 101002551) and by the Fonds de la Recherche Scientifique
--- FNRS under Grant No. F.4544.21. 
\end{acknowledgments}

\bibliography{references}

\begin{thebibliography}{103}%
\makeatletter
\providecommand \@ifxundefined [1]{%
 \@ifx{#1\undefined}
}%
\providecommand \@ifnum [1]{%
 \ifnum #1\expandafter \@firstoftwo
 \else \expandafter \@secondoftwo
 \fi
}%
\providecommand \@ifx [1]{%
 \ifx #1\expandafter \@firstoftwo
 \else \expandafter \@secondoftwo
 \fi
}%
\providecommand \natexlab [1]{#1}%
\providecommand \enquote  [1]{``#1''}%
\providecommand \bibnamefont  [1]{#1}%
\providecommand \bibfnamefont [1]{#1}%
\providecommand \citenamefont [1]{#1}%
\providecommand \href@noop [0]{\@secondoftwo}%
\providecommand \href [0]{\begingroup \@sanitize@url \@href}%
\providecommand \@href[1]{\@@startlink{#1}\@@href}%
\providecommand \@@href[1]{\endgroup#1\@@endlink}%
\providecommand \@sanitize@url [0]{\catcode `\\12\catcode `\$12\catcode
  `\&12\catcode `\#12\catcode `\^12\catcode `\_12\catcode `\%12\relax}%
\providecommand \@@startlink[1]{}%
\providecommand \@@endlink[0]{}%
\providecommand \url  [0]{\begingroup\@sanitize@url \@url }%
\providecommand \@url [1]{\endgroup\@href {#1}{\urlprefix }}%
\providecommand \urlprefix  [0]{URL }%
\providecommand \Eprint [0]{\href }%
\providecommand \doibase [0]{http://dx.doi.org/}%
\providecommand \selectlanguage [0]{\@gobble}%
\providecommand \bibinfo  [0]{\@secondoftwo}%
\providecommand \bibfield  [0]{\@secondoftwo}%
\providecommand \translation [1]{[#1]}%
\providecommand \BibitemOpen [0]{}%
\providecommand \bibitemStop [0]{}%
\providecommand \bibitemNoStop [0]{.\EOS\space}%
\providecommand \EOS [0]{\spacefactor3000\relax}%
\providecommand \BibitemShut  [1]{\csname bibitem#1\endcsname}%
\let\auto@bib@innerbib\@empty
\bibitem [{\citenamefont {Bekaert}\ \emph {et~al.}(2022)\citenamefont
  {Bekaert}, \citenamefont {Boulanger}, \citenamefont {Campoleoni},
  \citenamefont {Chiodaroli}, \citenamefont {Francia}, \citenamefont
  {Grigoriev}, \citenamefont {Sezgin},\ and\ \citenamefont
  {Skvortsov}}]{Bekaert:2022poo}%
  \BibitemOpen
  \bibfield  {author} {\bibinfo {author} {\bibfnamefont {X.}~\bibnamefont
  {Bekaert}}, \bibinfo {author} {\bibfnamefont {N.}~\bibnamefont {Boulanger}},
  \bibinfo {author} {\bibfnamefont {A.}~\bibnamefont {Campoleoni}}, \bibinfo
  {author} {\bibfnamefont {M.}~\bibnamefont {Chiodaroli}}, \bibinfo {author}
  {\bibfnamefont {D.}~\bibnamefont {Francia}}, \bibinfo {author} {\bibfnamefont
  {M.}~\bibnamefont {Grigoriev}}, \bibinfo {author} {\bibfnamefont
  {E.}~\bibnamefont {Sezgin}}, \ and\ \bibinfo {author} {\bibfnamefont
  {E.}~\bibnamefont {Skvortsov}},\ }\href@noop {} {\  (\bibinfo {year}
  {2022})},\ \Eprint {http://arxiv.org/abs/2205.01567} {arXiv:2205.01567
  [hep-th]} \BibitemShut {NoStop}%
\bibitem [{\citenamefont {Ponomarev}(2022)}]{Ponomarev:2022vjb}%
  \BibitemOpen
  \bibfield  {author} {\bibinfo {author} {\bibfnamefont {D.}~\bibnamefont
  {Ponomarev}},\ }\href@noop {} {\  (\bibinfo {year} {2022})},\ \Eprint
  {http://arxiv.org/abs/2206.15385} {arXiv:2206.15385 [hep-th]} \BibitemShut
  {NoStop}%
\bibitem [{\citenamefont {Fierz}\ and\ \citenamefont
  {Pauli}(1939)}]{Fierz:1939ix}%
  \BibitemOpen
  \bibfield  {author} {\bibinfo {author} {\bibfnamefont {M.}~\bibnamefont
  {Fierz}}\ and\ \bibinfo {author} {\bibfnamefont {W.}~\bibnamefont {Pauli}},\
  }\href {\doibase 10.1098/rspa.1939.0140} {\bibfield  {journal} {\bibinfo
  {journal} {Proc. Roy. Soc. Lond. A}\ }\textbf {\bibinfo {volume} {173}},\
  \bibinfo {pages} {211} (\bibinfo {year} {1939})}\BibitemShut {NoStop}%
\bibitem [{\citenamefont {Singh}\ and\ \citenamefont
  {Hagen}(1974{\natexlab{a}})}]{Singh:1974qz}%
  \BibitemOpen
  \bibfield  {author} {\bibinfo {author} {\bibfnamefont {L.~P.~S.}\
  \bibnamefont {Singh}}\ and\ \bibinfo {author} {\bibfnamefont {C.~R.}\
  \bibnamefont {Hagen}},\ }\href {\doibase 10.1103/PhysRevD.9.898} {\bibfield
  {journal} {\bibinfo  {journal} {Phys. Rev.}\ }\textbf {\bibinfo {volume}
  {D9}},\ \bibinfo {pages} {898} (\bibinfo {year}
  {1974}{\natexlab{a}})}\BibitemShut {NoStop}%
\bibitem [{\citenamefont {Singh}\ and\ \citenamefont
  {Hagen}(1974{\natexlab{b}})}]{Singh:1974rc}%
  \BibitemOpen
  \bibfield  {author} {\bibinfo {author} {\bibfnamefont {L.~P.~S.}\
  \bibnamefont {Singh}}\ and\ \bibinfo {author} {\bibfnamefont {C.~R.}\
  \bibnamefont {Hagen}},\ }\href {\doibase 10.1103/PhysRevD.9.910} {\bibfield
  {journal} {\bibinfo  {journal} {Phys. Rev.}\ }\textbf {\bibinfo {volume}
  {D9}},\ \bibinfo {pages} {910} (\bibinfo {year}
  {1974}{\natexlab{b}})}\BibitemShut {NoStop}%
\bibitem [{\citenamefont {Zinoviev}(2001)}]{Zinoviev:2001dt}%
  \BibitemOpen
  \bibfield  {author} {\bibinfo {author} {\bibfnamefont {Y.~M.}\ \bibnamefont
  {Zinoviev}},\ }\href@noop {} {\  (\bibinfo {year} {2001})},\ \Eprint
  {http://arxiv.org/abs/hep-th/0108192} {arXiv:hep-th/0108192} \BibitemShut
  {NoStop}%
\bibitem [{\citenamefont {Johnson}\ and\ \citenamefont
  {Sudarshan}(1961)}]{Johnson:1960vt}%
  \BibitemOpen
  \bibfield  {author} {\bibinfo {author} {\bibfnamefont {K.}~\bibnamefont
  {Johnson}}\ and\ \bibinfo {author} {\bibfnamefont {E.~C.~G.}\ \bibnamefont
  {Sudarshan}},\ }\href {\doibase 10.1016/0003-4916(61)90030-6} {\bibfield
  {journal} {\bibinfo  {journal} {Annals Phys.}\ }\textbf {\bibinfo {volume}
  {13}},\ \bibinfo {pages} {126} (\bibinfo {year} {1961})}\BibitemShut
  {NoStop}%
\bibitem [{\citenamefont {Velo}\ and\ \citenamefont
  {Zwanziger}(1969)}]{Velo:1969bt}%
  \BibitemOpen
  \bibfield  {author} {\bibinfo {author} {\bibfnamefont {G.}~\bibnamefont
  {Velo}}\ and\ \bibinfo {author} {\bibfnamefont {D.}~\bibnamefont
  {Zwanziger}},\ }\href {\doibase 10.1103/PhysRev.186.1337} {\bibfield
  {journal} {\bibinfo  {journal} {Phys. Rev.}\ }\textbf {\bibinfo {volume}
  {186}},\ \bibinfo {pages} {1337} (\bibinfo {year} {1969})}\BibitemShut
  {NoStop}%
\bibitem [{\citenamefont {Workman}\ \emph {et~al.}(2022)\citenamefont {Workman}
  \emph {et~al.}}]{ParticleDataGroup:2022pth}%
  \BibitemOpen
  \bibfield  {author} {\bibinfo {author} {\bibfnamefont {R.~L.}\ \bibnamefont
  {Workman}} \emph {et~al.} (\bibinfo {collaboration} {Particle Data Group}),\
  }\href {\doibase 10.1093/ptep/ptac097} {\bibfield  {journal} {\bibinfo
  {journal} {PTEP}\ }\textbf {\bibinfo {volume} {2022}},\ \bibinfo {pages}
  {083C01} (\bibinfo {year} {2022})}\BibitemShut {NoStop}%
\bibitem [{\citenamefont {Arkani-Hamed}\ \emph {et~al.}(2021)\citenamefont
  {Arkani-Hamed}, \citenamefont {Huang},\ and\ \citenamefont
  {Huang}}]{Arkani-Hamed:2017jhn}%
  \BibitemOpen
  \bibfield  {author} {\bibinfo {author} {\bibfnamefont {N.}~\bibnamefont
  {Arkani-Hamed}}, \bibinfo {author} {\bibfnamefont {T.-C.}\ \bibnamefont
  {Huang}}, \ and\ \bibinfo {author} {\bibfnamefont {Y.-t.}\ \bibnamefont
  {Huang}},\ }\href {\doibase 10.1007/JHEP11(2021)070} {\bibfield  {journal}
  {\bibinfo  {journal} {JHEP}\ }\textbf {\bibinfo {volume} {11}},\ \bibinfo
  {pages} {070} (\bibinfo {year} {2021})},\ \Eprint
  {http://arxiv.org/abs/1709.04891} {arXiv:1709.04891 [hep-th]} \BibitemShut
  {NoStop}%
\bibitem [{\citenamefont {Aoude}\ \emph {et~al.}(2020)\citenamefont {Aoude},
  \citenamefont {Haddad},\ and\ \citenamefont {Helset}}]{Aoude:2020onz}%
  \BibitemOpen
  \bibfield  {author} {\bibinfo {author} {\bibfnamefont {R.}~\bibnamefont
  {Aoude}}, \bibinfo {author} {\bibfnamefont {K.}~\bibnamefont {Haddad}}, \
  and\ \bibinfo {author} {\bibfnamefont {A.}~\bibnamefont {Helset}},\ }\href
  {\doibase 10.1007/JHEP05(2020)051} {\bibfield  {journal} {\bibinfo  {journal}
  {JHEP}\ }\textbf {\bibinfo {volume} {05}},\ \bibinfo {pages} {051} (\bibinfo
  {year} {2020})},\ \Eprint {http://arxiv.org/abs/2001.09164} {arXiv:2001.09164
  [hep-th]} \BibitemShut {NoStop}%
\bibitem [{\citenamefont {Lazopoulos}\ \emph {et~al.}(2022)\citenamefont
  {Lazopoulos}, \citenamefont {Ochirov},\ and\ \citenamefont
  {Shi}}]{Lazopoulos:2021mna}%
  \BibitemOpen
  \bibfield  {author} {\bibinfo {author} {\bibfnamefont {A.}~\bibnamefont
  {Lazopoulos}}, \bibinfo {author} {\bibfnamefont {A.}~\bibnamefont {Ochirov}},
  \ and\ \bibinfo {author} {\bibfnamefont {C.}~\bibnamefont {Shi}},\ }\href
  {\doibase 10.1007/JHEP03(2022)009} {\bibfield  {journal} {\bibinfo  {journal}
  {JHEP}\ }\textbf {\bibinfo {volume} {03}},\ \bibinfo {pages} {009} (\bibinfo
  {year} {2022})},\ \Eprint {http://arxiv.org/abs/2111.06847} {arXiv:2111.06847
  [hep-th]} \BibitemShut {NoStop}%
\bibitem [{\citenamefont {Britto}\ \emph
  {et~al.}(2005{\natexlab{a}})\citenamefont {Britto}, \citenamefont {Cachazo},\
  and\ \citenamefont {Feng}}]{Britto:2004ap}%
  \BibitemOpen
  \bibfield  {author} {\bibinfo {author} {\bibfnamefont {R.}~\bibnamefont
  {Britto}}, \bibinfo {author} {\bibfnamefont {F.}~\bibnamefont {Cachazo}}, \
  and\ \bibinfo {author} {\bibfnamefont {B.}~\bibnamefont {Feng}},\ }\href
  {\doibase 10.1016/j.nuclphysb.2005.02.030} {\bibfield  {journal} {\bibinfo
  {journal} {Nucl.Phys.}\ }\textbf {\bibinfo {volume} {B715}},\ \bibinfo
  {pages} {499} (\bibinfo {year} {2005}{\natexlab{a}})},\ \Eprint
  {http://arxiv.org/abs/hep-th/0412308} {arXiv:hep-th/0412308 [hep-th]}
  \BibitemShut {NoStop}%
\bibitem [{\citenamefont {Britto}\ \emph
  {et~al.}(2005{\natexlab{b}})\citenamefont {Britto}, \citenamefont {Cachazo},
  \citenamefont {Feng},\ and\ \citenamefont {Witten}}]{Britto:2005fq}%
  \BibitemOpen
  \bibfield  {author} {\bibinfo {author} {\bibfnamefont {R.}~\bibnamefont
  {Britto}}, \bibinfo {author} {\bibfnamefont {F.}~\bibnamefont {Cachazo}},
  \bibinfo {author} {\bibfnamefont {B.}~\bibnamefont {Feng}}, \ and\ \bibinfo
  {author} {\bibfnamefont {E.}~\bibnamefont {Witten}},\ }\href {\doibase
  10.1103/PhysRevLett.94.181602} {\bibfield  {journal} {\bibinfo  {journal}
  {Phys.Rev.Lett.}\ }\textbf {\bibinfo {volume} {94}},\ \bibinfo {pages}
  {181602} (\bibinfo {year} {2005}{\natexlab{b}})},\ \Eprint
  {http://arxiv.org/abs/hep-th/0501052} {arXiv:hep-th/0501052 [hep-th]}
  \BibitemShut {NoStop}%
\bibitem [{\citenamefont {Ochirov}(2018)}]{Ochirov:2018uyq}%
  \BibitemOpen
  \bibfield  {author} {\bibinfo {author} {\bibfnamefont {A.}~\bibnamefont
  {Ochirov}},\ }\href {\doibase 10.1007/JHEP04(2018)089} {\bibfield  {journal}
  {\bibinfo  {journal} {JHEP}\ }\textbf {\bibinfo {volume} {04}},\ \bibinfo
  {pages} {089} (\bibinfo {year} {2018})},\ \Eprint
  {http://arxiv.org/abs/1802.06730} {arXiv:1802.06730 [hep-ph]} \BibitemShut
  {NoStop}%
\bibitem [{\citenamefont {Johansson}\ and\ \citenamefont
  {Ochirov}(2019)}]{Johansson:2019dnu}%
  \BibitemOpen
  \bibfield  {author} {\bibinfo {author} {\bibfnamefont {H.}~\bibnamefont
  {Johansson}}\ and\ \bibinfo {author} {\bibfnamefont {A.}~\bibnamefont
  {Ochirov}},\ }\href {\doibase 10.1007/JHEP09(2019)040} {\bibfield  {journal}
  {\bibinfo  {journal} {JHEP}\ }\textbf {\bibinfo {volume} {09}},\ \bibinfo
  {pages} {040} (\bibinfo {year} {2019})},\ \Eprint
  {http://arxiv.org/abs/1906.12292} {arXiv:1906.12292 [hep-th]} \BibitemShut
  {NoStop}%
\bibitem [{\citenamefont {Badger}\ \emph {et~al.}(2005)\citenamefont {Badger},
  \citenamefont {Glover}, \citenamefont {Khoze},\ and\ \citenamefont
  {Svrcek}}]{Badger:2005zh}%
  \BibitemOpen
  \bibfield  {author} {\bibinfo {author} {\bibfnamefont {S.~D.}\ \bibnamefont
  {Badger}}, \bibinfo {author} {\bibfnamefont {E.~W.~N.}\ \bibnamefont
  {Glover}}, \bibinfo {author} {\bibfnamefont {V.~V.}\ \bibnamefont {Khoze}}, \
  and\ \bibinfo {author} {\bibfnamefont {P.}~\bibnamefont {Svrcek}},\ }\href
  {\doibase 10.1088/1126-6708/2005/07/025} {\bibfield  {journal} {\bibinfo
  {journal} {JHEP}\ }\textbf {\bibinfo {volume} {07}},\ \bibinfo {pages} {025}
  (\bibinfo {year} {2005})},\ \Eprint {http://arxiv.org/abs/hep-th/0504159}
  {arXiv:hep-th/0504159 [hep-th]} \BibitemShut {NoStop}%
\bibitem [{\citenamefont {Aoude}\ and\ \citenamefont
  {Machado}(2019)}]{Aoude:2019tzn}%
  \BibitemOpen
  \bibfield  {author} {\bibinfo {author} {\bibfnamefont {R.}~\bibnamefont
  {Aoude}}\ and\ \bibinfo {author} {\bibfnamefont {C.~S.}\ \bibnamefont
  {Machado}},\ }\href {\doibase 10.1007/JHEP12(2019)058} {\bibfield  {journal}
  {\bibinfo  {journal} {JHEP}\ }\textbf {\bibinfo {volume} {12}},\ \bibinfo
  {pages} {058} (\bibinfo {year} {2019})},\ \Eprint
  {http://arxiv.org/abs/1905.11433} {arXiv:1905.11433 [hep-ph]} \BibitemShut
  {NoStop}%
\bibitem [{\citenamefont {Metsaev}(1991{\natexlab{a}})}]{Metsaev:1991mt}%
  \BibitemOpen
  \bibfield  {author} {\bibinfo {author} {\bibfnamefont {R.~R.}\ \bibnamefont
  {Metsaev}},\ }\href {\doibase 10.1142/S0217732391000348} {\bibfield
  {journal} {\bibinfo  {journal} {Mod. Phys. Lett. A}\ }\textbf {\bibinfo
  {volume} {6}},\ \bibinfo {pages} {359} (\bibinfo {year}
  {1991}{\natexlab{a}})}\BibitemShut {NoStop}%
\bibitem [{\citenamefont {Metsaev}(1991{\natexlab{b}})}]{Metsaev:1991nb}%
  \BibitemOpen
  \bibfield  {author} {\bibinfo {author} {\bibfnamefont {R.~R.}\ \bibnamefont
  {Metsaev}},\ }\href {\doibase 10.1142/S0217732391002839} {\bibfield
  {journal} {\bibinfo  {journal} {Mod. Phys. Lett. A}\ }\textbf {\bibinfo
  {volume} {6}},\ \bibinfo {pages} {2411} (\bibinfo {year}
  {1991}{\natexlab{b}})}\BibitemShut {NoStop}%
\bibitem [{\citenamefont {Ponomarev}\ and\ \citenamefont
  {Skvortsov}(2017)}]{Ponomarev:2016lrm}%
  \BibitemOpen
  \bibfield  {author} {\bibinfo {author} {\bibfnamefont {D.}~\bibnamefont
  {Ponomarev}}\ and\ \bibinfo {author} {\bibfnamefont {E.~D.}\ \bibnamefont
  {Skvortsov}},\ }\href {\doibase 10.1088/1751-8121/aa56e7} {\bibfield
  {journal} {\bibinfo  {journal} {J. Phys. A}\ }\textbf {\bibinfo {volume}
  {50}},\ \bibinfo {pages} {095401} (\bibinfo {year} {2017})},\ \Eprint
  {http://arxiv.org/abs/1609.04655} {arXiv:1609.04655 [hep-th]} \BibitemShut
  {NoStop}%
\bibitem [{\citenamefont {Skvortsov}\ \emph {et~al.}(2018)\citenamefont
  {Skvortsov}, \citenamefont {Tran},\ and\ \citenamefont
  {Tsulaia}}]{Skvortsov:2018jea}%
  \BibitemOpen
  \bibfield  {author} {\bibinfo {author} {\bibfnamefont {E.~D.}\ \bibnamefont
  {Skvortsov}}, \bibinfo {author} {\bibfnamefont {T.}~\bibnamefont {Tran}}, \
  and\ \bibinfo {author} {\bibfnamefont {M.}~\bibnamefont {Tsulaia}},\ }\href
  {\doibase 10.1103/PhysRevLett.121.031601} {\bibfield  {journal} {\bibinfo
  {journal} {Phys. Rev. Lett.}\ }\textbf {\bibinfo {volume} {121}},\ \bibinfo
  {pages} {031601} (\bibinfo {year} {2018})},\ \Eprint
  {http://arxiv.org/abs/1805.00048} {arXiv:1805.00048 [hep-th]} \BibitemShut
  {NoStop}%
\bibitem [{\citenamefont {Krasnov}\ \emph {et~al.}(2021)\citenamefont
  {Krasnov}, \citenamefont {Skvortsov},\ and\ \citenamefont
  {Tran}}]{Krasnov:2021nsq}%
  \BibitemOpen
  \bibfield  {author} {\bibinfo {author} {\bibfnamefont {K.}~\bibnamefont
  {Krasnov}}, \bibinfo {author} {\bibfnamefont {E.}~\bibnamefont {Skvortsov}},
  \ and\ \bibinfo {author} {\bibfnamefont {T.}~\bibnamefont {Tran}},\ }\href
  {\doibase 10.1007/JHEP08(2021)076} {\bibfield  {journal} {\bibinfo  {journal}
  {JHEP}\ }\textbf {\bibinfo {volume} {08}},\ \bibinfo {pages} {076} (\bibinfo
  {year} {2021})},\ \Eprint {http://arxiv.org/abs/2105.12782} {arXiv:2105.12782
  [hep-th]} \BibitemShut {NoStop}%
\bibitem [{\citenamefont {Skvortsov}\ \emph {et~al.}(2020)\citenamefont
  {Skvortsov}, \citenamefont {Tran},\ and\ \citenamefont
  {Tsulaia}}]{Skvortsov:2020wtf}%
  \BibitemOpen
  \bibfield  {author} {\bibinfo {author} {\bibfnamefont {E.}~\bibnamefont
  {Skvortsov}}, \bibinfo {author} {\bibfnamefont {T.}~\bibnamefont {Tran}}, \
  and\ \bibinfo {author} {\bibfnamefont {M.}~\bibnamefont {Tsulaia}},\ }\href
  {\doibase 10.1103/PhysRevD.101.106001} {\bibfield  {journal} {\bibinfo
  {journal} {Phys. Rev. D}\ }\textbf {\bibinfo {volume} {101}},\ \bibinfo
  {pages} {106001} (\bibinfo {year} {2020})},\ \Eprint
  {http://arxiv.org/abs/2002.08487} {arXiv:2002.08487 [hep-th]} \BibitemShut
  {NoStop}%
\bibitem [{\citenamefont {Skvortsov}\ and\ \citenamefont
  {Tran}(2020)}]{Skvortsov:2020gpn}%
  \BibitemOpen
  \bibfield  {author} {\bibinfo {author} {\bibfnamefont {E.}~\bibnamefont
  {Skvortsov}}\ and\ \bibinfo {author} {\bibfnamefont {T.}~\bibnamefont
  {Tran}},\ }\href {\doibase 10.1007/JHEP07(2020)021} {\bibfield  {journal}
  {\bibinfo  {journal} {JHEP}\ }\textbf {\bibinfo {volume} {07}},\ \bibinfo
  {pages} {021} (\bibinfo {year} {2020})},\ \Eprint
  {http://arxiv.org/abs/2004.10797} {arXiv:2004.10797 [hep-th]} \BibitemShut
  {NoStop}%
\bibitem [{\citenamefont {Sharapov}\ \emph {et~al.}(2022)\citenamefont
  {Sharapov}, \citenamefont {Skvortsov}, \citenamefont {Sukhanov},\ and\
  \citenamefont {Van~Dongen}}]{Sharapov:2022faa}%
  \BibitemOpen
  \bibfield  {author} {\bibinfo {author} {\bibfnamefont {A.}~\bibnamefont
  {Sharapov}}, \bibinfo {author} {\bibfnamefont {E.}~\bibnamefont {Skvortsov}},
  \bibinfo {author} {\bibfnamefont {A.}~\bibnamefont {Sukhanov}}, \ and\
  \bibinfo {author} {\bibfnamefont {R.}~\bibnamefont {Van~Dongen}},\
  }\href@noop {} {\  (\bibinfo {year} {2022})},\ \Eprint
  {http://arxiv.org/abs/2205.07794} {arXiv:2205.07794 [hep-th]} \BibitemShut
  {NoStop}%
\bibitem [{\citenamefont {Sharapov}\ and\ \citenamefont
  {Skvortsov}(2022)}]{Sharapov:2022awp}%
  \BibitemOpen
  \bibfield  {author} {\bibinfo {author} {\bibfnamefont {A.}~\bibnamefont
  {Sharapov}}\ and\ \bibinfo {author} {\bibfnamefont {E.}~\bibnamefont
  {Skvortsov}},\ }\href@noop {} {\  (\bibinfo {year} {2022})},\ \Eprint
  {http://arxiv.org/abs/2205.15293} {arXiv:2205.15293 [hep-th]} \BibitemShut
  {NoStop}%
\bibitem [{\citenamefont {Conde}\ and\ \citenamefont
  {Marzolla}(2016)}]{Conde:2016vxs}%
  \BibitemOpen
  \bibfield  {author} {\bibinfo {author} {\bibfnamefont {E.}~\bibnamefont
  {Conde}}\ and\ \bibinfo {author} {\bibfnamefont {A.}~\bibnamefont
  {Marzolla}},\ }\href {\doibase 10.1007/JHEP09(2016)041} {\bibfield  {journal}
  {\bibinfo  {journal} {JHEP}\ }\textbf {\bibinfo {volume} {09}},\ \bibinfo
  {pages} {041} (\bibinfo {year} {2016})},\ \Eprint
  {http://arxiv.org/abs/1601.08113} {arXiv:1601.08113 [hep-th]} \BibitemShut
  {NoStop}%
\bibitem [{\citenamefont {Conde}\ \emph {et~al.}(2016)\citenamefont {Conde},
  \citenamefont {Joung},\ and\ \citenamefont {Mkrtchyan}}]{Conde:2016izb}%
  \BibitemOpen
  \bibfield  {author} {\bibinfo {author} {\bibfnamefont {E.}~\bibnamefont
  {Conde}}, \bibinfo {author} {\bibfnamefont {E.}~\bibnamefont {Joung}}, \ and\
  \bibinfo {author} {\bibfnamefont {K.}~\bibnamefont {Mkrtchyan}},\ }\href
  {\doibase 10.1007/JHEP08(2016)040} {\bibfield  {journal} {\bibinfo  {journal}
  {JHEP}\ }\textbf {\bibinfo {volume} {08}},\ \bibinfo {pages} {040} (\bibinfo
  {year} {2016})},\ \Eprint {http://arxiv.org/abs/1605.07402} {arXiv:1605.07402
  [hep-th]} \BibitemShut {NoStop}%
\bibitem [{\citenamefont {Guevara}\ \emph
  {et~al.}(2019{\natexlab{a}})\citenamefont {Guevara}, \citenamefont
  {Ochirov},\ and\ \citenamefont {Vines}}]{Guevara:2018wpp}%
  \BibitemOpen
  \bibfield  {author} {\bibinfo {author} {\bibfnamefont {A.}~\bibnamefont
  {Guevara}}, \bibinfo {author} {\bibfnamefont {A.}~\bibnamefont {Ochirov}}, \
  and\ \bibinfo {author} {\bibfnamefont {J.}~\bibnamefont {Vines}},\ }\href
  {\doibase 10.1007/JHEP09(2019)056} {\bibfield  {journal} {\bibinfo  {journal}
  {JHEP}\ }\textbf {\bibinfo {volume} {09}},\ \bibinfo {pages} {056} (\bibinfo
  {year} {2019}{\natexlab{a}})},\ \Eprint {http://arxiv.org/abs/1812.06895}
  {arXiv:1812.06895 [hep-th]} \BibitemShut {NoStop}%
\bibitem [{\citenamefont {Chung}\ \emph {et~al.}(2019)\citenamefont {Chung},
  \citenamefont {Huang}, \citenamefont {Kim},\ and\ \citenamefont
  {Lee}}]{Chung:2018kqs}%
  \BibitemOpen
  \bibfield  {author} {\bibinfo {author} {\bibfnamefont {M.-Z.}\ \bibnamefont
  {Chung}}, \bibinfo {author} {\bibfnamefont {Y.-T.}\ \bibnamefont {Huang}},
  \bibinfo {author} {\bibfnamefont {J.-W.}\ \bibnamefont {Kim}}, \ and\
  \bibinfo {author} {\bibfnamefont {S.}~\bibnamefont {Lee}},\ }\href {\doibase
  10.1007/JHEP04(2019)156} {\bibfield  {journal} {\bibinfo  {journal} {JHEP}\
  }\textbf {\bibinfo {volume} {04}},\ \bibinfo {pages} {156} (\bibinfo {year}
  {2019})},\ \Eprint {http://arxiv.org/abs/1812.08752} {arXiv:1812.08752
  [hep-th]} \BibitemShut {NoStop}%
\bibitem [{\citenamefont {Ward}(1977)}]{Ward:1977ta}%
  \BibitemOpen
  \bibfield  {author} {\bibinfo {author} {\bibfnamefont {R.~S.}\ \bibnamefont
  {Ward}},\ }\href {\doibase 10.1016/0375-9601(77)90842-8} {\bibfield
  {journal} {\bibinfo  {journal} {Phys. Lett.}\ }\textbf {\bibinfo {volume}
  {A61}},\ \bibinfo {pages} {81} (\bibinfo {year} {1977})}\BibitemShut
  {NoStop}%
\bibitem [{\citenamefont {Parkes}(1992)}]{Parkes:1992rz}%
  \BibitemOpen
  \bibfield  {author} {\bibinfo {author} {\bibfnamefont {A.}~\bibnamefont
  {Parkes}},\ }\href {\doibase 10.1016/0370-2693(92)91773-3} {\bibfield
  {journal} {\bibinfo  {journal} {Phys. Lett. B}\ }\textbf {\bibinfo {volume}
  {286}},\ \bibinfo {pages} {265} (\bibinfo {year} {1992})},\ \Eprint
  {http://arxiv.org/abs/hep-th/9203074} {arXiv:hep-th/9203074} \BibitemShut
  {NoStop}%
\bibitem [{\citenamefont {Siegel}(1992)}]{Siegel:1992xp}%
  \BibitemOpen
  \bibfield  {author} {\bibinfo {author} {\bibfnamefont {W.}~\bibnamefont
  {Siegel}},\ }\href {\doibase 10.1103/PhysRevD.46.R3235} {\bibfield  {journal}
  {\bibinfo  {journal} {Phys. Rev. D}\ }\textbf {\bibinfo {volume} {46}},\
  \bibinfo {pages} {R3235} (\bibinfo {year} {1992})},\ \Eprint
  {http://arxiv.org/abs/hep-th/9205075} {arXiv:hep-th/9205075} \BibitemShut
  {NoStop}%
\bibitem [{\citenamefont {Chalmers}\ and\ \citenamefont
  {Siegel}(1996)}]{Chalmers:1996rq}%
  \BibitemOpen
  \bibfield  {author} {\bibinfo {author} {\bibfnamefont {G.}~\bibnamefont
  {Chalmers}}\ and\ \bibinfo {author} {\bibfnamefont {W.}~\bibnamefont
  {Siegel}},\ }\href {\doibase 10.1103/PhysRevD.54.7628} {\bibfield  {journal}
  {\bibinfo  {journal} {Phys. Rev. D}\ }\textbf {\bibinfo {volume} {54}},\
  \bibinfo {pages} {7628} (\bibinfo {year} {1996})},\ \Eprint
  {http://arxiv.org/abs/hep-th/9606061} {arXiv:hep-th/9606061} \BibitemShut
  {NoStop}%
\bibitem [{\citenamefont {Rosly}\ and\ \citenamefont
  {Selivanov}(1997)}]{Rosly:1996vr}%
  \BibitemOpen
  \bibfield  {author} {\bibinfo {author} {\bibfnamefont {A.~A.}\ \bibnamefont
  {Rosly}}\ and\ \bibinfo {author} {\bibfnamefont {K.~G.}\ \bibnamefont
  {Selivanov}},\ }\href {\doibase 10.1016/S0370-2693(97)00268-2} {\bibfield
  {journal} {\bibinfo  {journal} {Phys. Lett. B}\ }\textbf {\bibinfo {volume}
  {399}},\ \bibinfo {pages} {135} (\bibinfo {year} {1997})},\ \Eprint
  {http://arxiv.org/abs/hep-th/9611101} {arXiv:hep-th/9611101} \BibitemShut
  {NoStop}%
\bibitem [{\citenamefont {Abou-Zeid}\ and\ \citenamefont
  {Hull}(2006)}]{Abou-Zeid:2005zfo}%
  \BibitemOpen
  \bibfield  {author} {\bibinfo {author} {\bibfnamefont {M.}~\bibnamefont
  {Abou-Zeid}}\ and\ \bibinfo {author} {\bibfnamefont {C.~M.}\ \bibnamefont
  {Hull}},\ }\href {\doibase 10.1088/1126-6708/2006/02/057} {\bibfield
  {journal} {\bibinfo  {journal} {JHEP}\ }\textbf {\bibinfo {volume} {02}},\
  \bibinfo {pages} {057} (\bibinfo {year} {2006})},\ \Eprint
  {http://arxiv.org/abs/hep-th/0511189} {arXiv:hep-th/0511189} \BibitemShut
  {NoStop}%
\bibitem [{\citenamefont {Adamo}\ \emph
  {et~al.}(2020{\natexlab{a}})\citenamefont {Adamo}, \citenamefont {Mason},\
  and\ \citenamefont {Sharma}}]{Adamo:2020syc}%
  \BibitemOpen
  \bibfield  {author} {\bibinfo {author} {\bibfnamefont {T.}~\bibnamefont
  {Adamo}}, \bibinfo {author} {\bibfnamefont {L.}~\bibnamefont {Mason}}, \ and\
  \bibinfo {author} {\bibfnamefont {A.}~\bibnamefont {Sharma}},\ }\href
  {\doibase 10.1103/PhysRevLett.125.041602} {\bibfield  {journal} {\bibinfo
  {journal} {Phys. Rev. Lett.}\ }\textbf {\bibinfo {volume} {125}},\ \bibinfo
  {pages} {041602} (\bibinfo {year} {2020}{\natexlab{a}})},\ \Eprint
  {http://arxiv.org/abs/2003.13501} {arXiv:2003.13501 [hep-th]} \BibitemShut
  {NoStop}%
\bibitem [{\citenamefont {Adamo}\ \emph
  {et~al.}(2020{\natexlab{b}})\citenamefont {Adamo}, \citenamefont {Mason},\
  and\ \citenamefont {Sharma}}]{Adamo:2020yzi}%
  \BibitemOpen
  \bibfield  {author} {\bibinfo {author} {\bibfnamefont {T.}~\bibnamefont
  {Adamo}}, \bibinfo {author} {\bibfnamefont {L.}~\bibnamefont {Mason}}, \ and\
  \bibinfo {author} {\bibfnamefont {A.}~\bibnamefont {Sharma}},\ }\href@noop {}
  {\  (\bibinfo {year} {2020}{\natexlab{b}})},\ \Eprint
  {http://arxiv.org/abs/2010.14996} {arXiv:2010.14996 [hep-th]} \BibitemShut
  {NoStop}%
\bibitem [{\citenamefont {Arkani-Hamed}\ and\ \citenamefont
  {Kaplan}(2008)}]{ArkaniHamed:2008yf}%
  \BibitemOpen
  \bibfield  {author} {\bibinfo {author} {\bibfnamefont {N.}~\bibnamefont
  {Arkani-Hamed}}\ and\ \bibinfo {author} {\bibfnamefont {J.}~\bibnamefont
  {Kaplan}},\ }\href {\doibase 10.1088/1126-6708/2008/04/076} {\bibfield
  {journal} {\bibinfo  {journal} {JHEP}\ }\textbf {\bibinfo {volume} {0804}},\
  \bibinfo {pages} {076} (\bibinfo {year} {2008})},\ \Eprint
  {http://arxiv.org/abs/0801.2385} {arXiv:0801.2385 [hep-th]} \BibitemShut
  {NoStop}%
\bibitem [{\citenamefont {Dixon}(1996)}]{Dixon:1996wi}%
  \BibitemOpen
  \bibfield  {author} {\bibinfo {author} {\bibfnamefont {L.~J.}\ \bibnamefont
  {Dixon}},\ }in\ \href
  {http://www-public.slac.stanford.edu/sciDoc/docMeta.aspx?slacPubNumber=SLAC-PUB-7106}
  {\emph {\bibinfo {booktitle} {{QCD and beyond. Proceedings, Theoretical
  Advanced Study Institute in Elementary Particle Physics, TASI-95, Boulder,
  USA, June 4-30, 1995}}}}\ (\bibinfo {year} {1996})\ pp.\ \bibinfo {pages}
  {539--584},\ \Eprint {http://arxiv.org/abs/hep-ph/9601359}
  {arXiv:hep-ph/9601359 [hep-ph]} \BibitemShut {NoStop}%
\bibitem [{\citenamefont {Ferrario}\ \emph {et~al.}(2006)\citenamefont
  {Ferrario}, \citenamefont {Rodrigo},\ and\ \citenamefont
  {Talavera}}]{Ferrario:2006np}%
  \BibitemOpen
  \bibfield  {author} {\bibinfo {author} {\bibfnamefont {P.}~\bibnamefont
  {Ferrario}}, \bibinfo {author} {\bibfnamefont {G.}~\bibnamefont {Rodrigo}}, \
  and\ \bibinfo {author} {\bibfnamefont {P.}~\bibnamefont {Talavera}},\ }\href
  {\doibase 10.1103/PhysRevLett.96.182001} {\bibfield  {journal} {\bibinfo
  {journal} {Phys.Rev.Lett.}\ }\textbf {\bibinfo {volume} {96}},\ \bibinfo
  {pages} {182001} (\bibinfo {year} {2006})},\ \Eprint
  {http://arxiv.org/abs/hep-th/0602043} {arXiv:hep-th/0602043 [hep-th]}
  \BibitemShut {NoStop}%
\bibitem [{\citenamefont {Schwinn}\ and\ \citenamefont
  {Weinzierl}(2007)}]{Schwinn:2007ee}%
  \BibitemOpen
  \bibfield  {author} {\bibinfo {author} {\bibfnamefont {C.}~\bibnamefont
  {Schwinn}}\ and\ \bibinfo {author} {\bibfnamefont {S.}~\bibnamefont
  {Weinzierl}},\ }\href {\doibase 10.1088/1126-6708/2007/04/072} {\bibfield
  {journal} {\bibinfo  {journal} {JHEP}\ }\textbf {\bibinfo {volume} {0704}},\
  \bibinfo {pages} {072} (\bibinfo {year} {2007})},\ \Eprint
  {http://arxiv.org/abs/hep-ph/0703021} {arXiv:hep-ph/0703021 [HEP-PH]}
  \BibitemShut {NoStop}%
\bibitem [{\citenamefont {Ballav}\ and\ \citenamefont
  {Manna}(2021)}]{Ballav:2021ahg}%
  \BibitemOpen
  \bibfield  {author} {\bibinfo {author} {\bibfnamefont {S.}~\bibnamefont
  {Ballav}}\ and\ \bibinfo {author} {\bibfnamefont {A.}~\bibnamefont {Manna}},\
  }\href@noop {} {\  (\bibinfo {year} {2021})},\ \Eprint
  {http://arxiv.org/abs/2109.06546} {arXiv:2109.06546 [hep-th]} \BibitemShut
  {NoStop}%
\bibitem [{\citenamefont {Badger}\ \emph {et~al.}(2009)\citenamefont {Badger},
  \citenamefont {Bjerrum-Bohr},\ and\ \citenamefont {Vanhove}}]{Badger:2008rn}%
  \BibitemOpen
  \bibfield  {author} {\bibinfo {author} {\bibfnamefont {S.}~\bibnamefont
  {Badger}}, \bibinfo {author} {\bibfnamefont {N.~E.~J.}\ \bibnamefont
  {Bjerrum-Bohr}}, \ and\ \bibinfo {author} {\bibfnamefont {P.}~\bibnamefont
  {Vanhove}},\ }\href {\doibase 10.1088/1126-6708/2009/02/038} {\bibfield
  {journal} {\bibinfo  {journal} {JHEP}\ }\textbf {\bibinfo {volume} {02}},\
  \bibinfo {pages} {038} (\bibinfo {year} {2009})},\ \Eprint
  {http://arxiv.org/abs/0811.3405} {arXiv:0811.3405 [hep-th]} \BibitemShut
  {NoStop}%
\bibitem [{\citenamefont {Penrose}(1976{\natexlab{a}})}]{Penrose:1976js}%
  \BibitemOpen
  \bibfield  {author} {\bibinfo {author} {\bibfnamefont {R.}~\bibnamefont
  {Penrose}},\ }\href {\doibase 10.1007/BF00762011} {\bibfield  {journal}
  {\bibinfo  {journal} {Gen. Rel. Grav.}\ }\textbf {\bibinfo {volume} {7}},\
  \bibinfo {pages} {31} (\bibinfo {year} {1976}{\natexlab{a}})}\BibitemShut
  {NoStop}%
\bibitem [{\citenamefont {Penrose}(1976{\natexlab{b}})}]{Penrose:1976jq}%
  \BibitemOpen
  \bibfield  {author} {\bibinfo {author} {\bibfnamefont {R.}~\bibnamefont
  {Penrose}},\ }\href {\doibase 10.1007/BF00763433} {\bibfield  {journal}
  {\bibinfo  {journal} {Gen. Rel. Grav.}\ }\textbf {\bibinfo {volume} {7}},\
  \bibinfo {pages} {171} (\bibinfo {year} {1976}{\natexlab{b}})}\BibitemShut
  {NoStop}%
\bibitem [{\citenamefont {Capovilla}\ \emph {et~al.}(1991)\citenamefont
  {Capovilla}, \citenamefont {Jacobson}, \citenamefont {Dell},\ and\
  \citenamefont {Mason}}]{Capovilla:1991qb}%
  \BibitemOpen
  \bibfield  {author} {\bibinfo {author} {\bibfnamefont {R.}~\bibnamefont
  {Capovilla}}, \bibinfo {author} {\bibfnamefont {T.}~\bibnamefont {Jacobson}},
  \bibinfo {author} {\bibfnamefont {J.}~\bibnamefont {Dell}}, \ and\ \bibinfo
  {author} {\bibfnamefont {L.~J.}\ \bibnamefont {Mason}},\ }\href {\doibase
  10.1088/0264-9381/8/1/009} {\bibfield  {journal} {\bibinfo  {journal} {Class.
  Quant. Grav.}\ }\textbf {\bibinfo {volume} {8}},\ \bibinfo {pages} {41}
  (\bibinfo {year} {1991})}\BibitemShut {NoStop}%
\bibitem [{\citenamefont {Landau}\ and\ \citenamefont
  {Lifshitz}(1975)}]{Landau:1982dva}%
  \BibitemOpen
  \bibfield  {author} {\bibinfo {author} {\bibfnamefont {L.~D.}\ \bibnamefont
  {Landau}}\ and\ \bibinfo {author} {\bibfnamefont {E.~M.}\ \bibnamefont
  {Lifshitz}},\ }\href@noop {} {\emph {\bibinfo {title} {{The Classical Theory
  of Fields}}}},\ \bibinfo {series} {Course of Theoretical Physics},
  Vol.~\bibinfo {volume} {2}\ (\bibinfo  {publisher} {Pergamon Press},\
  \bibinfo {address} {Oxford},\ \bibinfo {year} {1975})\BibitemShut {NoStop}%
\bibitem [{\citenamefont {Poisson}\ and\ \citenamefont
  {Will}(2014)}]{Poisson:2014}%
  \BibitemOpen
  \bibfield  {author} {\bibinfo {author} {\bibfnamefont {E.}~\bibnamefont
  {Poisson}}\ and\ \bibinfo {author} {\bibfnamefont {C.~M.}\ \bibnamefont
  {Will}},\ }\href@noop {} {\emph {\bibinfo {title} {{Gravity: Newtonian,
  Post-Newtonian, Relativistic}}}}\ (\bibinfo  {publisher} {Cambridge
  University Press},\ \bibinfo {year} {2014})\BibitemShut {NoStop}%
\bibitem [{\citenamefont {Kawai}\ \emph {et~al.}(1986)\citenamefont {Kawai},
  \citenamefont {Lewellen},\ and\ \citenamefont {Tye}}]{Kawai:1985xq}%
  \BibitemOpen
  \bibfield  {author} {\bibinfo {author} {\bibfnamefont {H.}~\bibnamefont
  {Kawai}}, \bibinfo {author} {\bibfnamefont {D.}~\bibnamefont {Lewellen}}, \
  and\ \bibinfo {author} {\bibfnamefont {S.}~\bibnamefont {Tye}},\ }\href
  {\doibase 10.1016/0550-3213(86)90362-7} {\bibfield  {journal} {\bibinfo
  {journal} {Nucl.Phys.}\ }\textbf {\bibinfo {volume} {B269}},\ \bibinfo
  {pages} {1} (\bibinfo {year} {1986})}\BibitemShut {NoStop}%
\bibitem [{\citenamefont {Bern}\ \emph {et~al.}(2008)\citenamefont {Bern},
  \citenamefont {Carrasco},\ and\ \citenamefont {Johansson}}]{Bern:2008qj}%
  \BibitemOpen
  \bibfield  {author} {\bibinfo {author} {\bibfnamefont {Z.}~\bibnamefont
  {Bern}}, \bibinfo {author} {\bibfnamefont {J.~J.~M.}\ \bibnamefont
  {Carrasco}}, \ and\ \bibinfo {author} {\bibfnamefont {H.}~\bibnamefont
  {Johansson}},\ }\href {\doibase 10.1103/PhysRevD.78.085011} {\bibfield
  {journal} {\bibinfo  {journal} {Phys.Rev.}\ }\textbf {\bibinfo {volume}
  {D78}},\ \bibinfo {pages} {085011} (\bibinfo {year} {2008})},\ \Eprint
  {http://arxiv.org/abs/0805.3993} {arXiv:0805.3993 [hep-ph]} \BibitemShut
  {NoStop}%
\bibitem [{\citenamefont {Bjerrum-Bohr}\ \emph {et~al.}(2014)\citenamefont
  {Bjerrum-Bohr}, \citenamefont {Donoghue},\ and\ \citenamefont
  {Vanhove}}]{Bjerrum-Bohr:2013bxa}%
  \BibitemOpen
  \bibfield  {author} {\bibinfo {author} {\bibfnamefont {N.~E.~J.}\
  \bibnamefont {Bjerrum-Bohr}}, \bibinfo {author} {\bibfnamefont {J.~F.}\
  \bibnamefont {Donoghue}}, \ and\ \bibinfo {author} {\bibfnamefont
  {P.}~\bibnamefont {Vanhove}},\ }\href {\doibase 10.1007/JHEP02(2014)111}
  {\bibfield  {journal} {\bibinfo  {journal} {JHEP}\ }\textbf {\bibinfo
  {volume} {02}},\ \bibinfo {pages} {111} (\bibinfo {year} {2014})},\ \Eprint
  {http://arxiv.org/abs/1309.0804} {arXiv:1309.0804 [hep-th]} \BibitemShut
  {NoStop}%
\bibitem [{\citenamefont {Naculich}(2014)}]{Naculich:2014naa}%
  \BibitemOpen
  \bibfield  {author} {\bibinfo {author} {\bibfnamefont {S.~G.}\ \bibnamefont
  {Naculich}},\ }\href {\doibase 10.1007/JHEP09(2014)029} {\bibfield  {journal}
  {\bibinfo  {journal} {JHEP}\ }\textbf {\bibinfo {volume} {1409}},\ \bibinfo
  {pages} {029} (\bibinfo {year} {2014})},\ \Eprint
  {http://arxiv.org/abs/1407.7836} {arXiv:1407.7836 [hep-th]} \BibitemShut
  {NoStop}%
\bibitem [{\citenamefont {Plefka}\ \emph {et~al.}(2020)\citenamefont {Plefka},
  \citenamefont {Shi},\ and\ \citenamefont {Wang}}]{Plefka:2019wyg}%
  \BibitemOpen
  \bibfield  {author} {\bibinfo {author} {\bibfnamefont {J.}~\bibnamefont
  {Plefka}}, \bibinfo {author} {\bibfnamefont {C.}~\bibnamefont {Shi}}, \ and\
  \bibinfo {author} {\bibfnamefont {T.}~\bibnamefont {Wang}},\ }\href {\doibase
  10.1103/PhysRevD.101.066004} {\bibfield  {journal} {\bibinfo  {journal}
  {Phys. Rev. D}\ }\textbf {\bibinfo {volume} {101}},\ \bibinfo {pages}
  {066004} (\bibinfo {year} {2020})},\ \Eprint
  {http://arxiv.org/abs/1911.06785} {arXiv:1911.06785 [hep-th]} \BibitemShut
  {NoStop}%
\bibitem [{\citenamefont {Johansson}\ and\ \citenamefont
  {Ochirov}(2016)}]{Johansson:2015oia}%
  \BibitemOpen
  \bibfield  {author} {\bibinfo {author} {\bibfnamefont {H.}~\bibnamefont
  {Johansson}}\ and\ \bibinfo {author} {\bibfnamefont {A.}~\bibnamefont
  {Ochirov}},\ }\href {\doibase 10.1007/JHEP01(2016)170} {\bibfield  {journal}
  {\bibinfo  {journal} {JHEP}\ }\textbf {\bibinfo {volume} {01}},\ \bibinfo
  {pages} {170} (\bibinfo {year} {2016})},\ \Eprint
  {http://arxiv.org/abs/1507.00332} {arXiv:1507.00332 [hep-ph]} \BibitemShut
  {NoStop}%
\bibitem [{\citenamefont {de~la Cruz}\ \emph {et~al.}(2016)\citenamefont {de~la
  Cruz}, \citenamefont {Kniss},\ and\ \citenamefont
  {Weinzierl}}]{delaCruz:2016wbr}%
  \BibitemOpen
  \bibfield  {author} {\bibinfo {author} {\bibfnamefont {L.}~\bibnamefont
  {de~la Cruz}}, \bibinfo {author} {\bibfnamefont {A.}~\bibnamefont {Kniss}}, \
  and\ \bibinfo {author} {\bibfnamefont {S.}~\bibnamefont {Weinzierl}},\ }\href
  {\doibase 10.1103/PhysRevLett.116.201601} {\bibfield  {journal} {\bibinfo
  {journal} {Phys. Rev. Lett.}\ }\textbf {\bibinfo {volume} {116}},\ \bibinfo
  {pages} {201601} (\bibinfo {year} {2016})},\ \Eprint
  {http://arxiv.org/abs/1601.04523} {arXiv:1601.04523 [hep-th]} \BibitemShut
  {NoStop}%
\bibitem [{\citenamefont {Brown}\ and\ \citenamefont
  {Naculich}(2018)}]{Brown:2018wss}%
  \BibitemOpen
  \bibfield  {author} {\bibinfo {author} {\bibfnamefont {R.~W.}\ \bibnamefont
  {Brown}}\ and\ \bibinfo {author} {\bibfnamefont {S.~G.}\ \bibnamefont
  {Naculich}},\ }\href {\doibase 10.1007/JHEP03(2018)057} {\bibfield  {journal}
  {\bibinfo  {journal} {JHEP}\ }\textbf {\bibinfo {volume} {03}},\ \bibinfo
  {pages} {057} (\bibinfo {year} {2018})},\ \Eprint
  {http://arxiv.org/abs/1802.01620} {arXiv:1802.01620 [hep-th]} \BibitemShut
  {NoStop}%
\bibitem [{\citenamefont {Bautista}\ and\ \citenamefont
  {Guevara}(2021)}]{Bautista:2019evw}%
  \BibitemOpen
  \bibfield  {author} {\bibinfo {author} {\bibfnamefont {Y.~F.}\ \bibnamefont
  {Bautista}}\ and\ \bibinfo {author} {\bibfnamefont {A.}~\bibnamefont
  {Guevara}},\ }\href {\doibase 10.1007/JHEP11(2021)184} {\bibfield  {journal}
  {\bibinfo  {journal} {JHEP}\ }\textbf {\bibinfo {volume} {11}},\ \bibinfo
  {pages} {184} (\bibinfo {year} {2021})},\ \Eprint
  {http://arxiv.org/abs/1908.11349} {arXiv:1908.11349 [hep-th]} \BibitemShut
  {NoStop}%
\bibitem [{\citenamefont {Guevara}(2019)}]{Guevara:2017csg}%
  \BibitemOpen
  \bibfield  {author} {\bibinfo {author} {\bibfnamefont {A.}~\bibnamefont
  {Guevara}},\ }\href {\doibase 10.1007/JHEP04(2019)033} {\bibfield  {journal}
  {\bibinfo  {journal} {JHEP}\ }\textbf {\bibinfo {volume} {04}},\ \bibinfo
  {pages} {033} (\bibinfo {year} {2019})},\ \Eprint
  {http://arxiv.org/abs/1706.02314} {arXiv:1706.02314 [hep-th]} \BibitemShut
  {NoStop}%
\bibitem [{\citenamefont {Maybee}\ \emph {et~al.}(2019)\citenamefont {Maybee},
  \citenamefont {O'Connell},\ and\ \citenamefont {Vines}}]{Maybee:2019jus}%
  \BibitemOpen
  \bibfield  {author} {\bibinfo {author} {\bibfnamefont {B.}~\bibnamefont
  {Maybee}}, \bibinfo {author} {\bibfnamefont {D.}~\bibnamefont {O'Connell}}, \
  and\ \bibinfo {author} {\bibfnamefont {J.}~\bibnamefont {Vines}},\ }\href
  {\doibase 10.1007/JHEP12(2019)156} {\bibfield  {journal} {\bibinfo  {journal}
  {JHEP}\ }\textbf {\bibinfo {volume} {12}},\ \bibinfo {pages} {156} (\bibinfo
  {year} {2019})},\ \Eprint {http://arxiv.org/abs/1906.09260} {arXiv:1906.09260
  [hep-th]} \BibitemShut {NoStop}%
\bibitem [{\citenamefont {Guevara}\ \emph
  {et~al.}(2019{\natexlab{b}})\citenamefont {Guevara}, \citenamefont
  {Ochirov},\ and\ \citenamefont {Vines}}]{Guevara:2019fsj}%
  \BibitemOpen
  \bibfield  {author} {\bibinfo {author} {\bibfnamefont {A.}~\bibnamefont
  {Guevara}}, \bibinfo {author} {\bibfnamefont {A.}~\bibnamefont {Ochirov}}, \
  and\ \bibinfo {author} {\bibfnamefont {J.}~\bibnamefont {Vines}},\ }\href
  {\doibase 10.1103/PhysRevD.100.104024} {\bibfield  {journal} {\bibinfo
  {journal} {Phys. Rev.}\ }\textbf {\bibinfo {volume} {D100}},\ \bibinfo
  {pages} {104024} (\bibinfo {year} {2019}{\natexlab{b}})},\ \Eprint
  {http://arxiv.org/abs/1906.10071} {arXiv:1906.10071 [hep-th]} \BibitemShut
  {NoStop}%
\bibitem [{\citenamefont {Arkani-Hamed}\ \emph {et~al.}(2020)\citenamefont
  {Arkani-Hamed}, \citenamefont {Huang},\ and\ \citenamefont
  {O'Connell}}]{Arkani-Hamed:2019ymq}%
  \BibitemOpen
  \bibfield  {author} {\bibinfo {author} {\bibfnamefont {N.}~\bibnamefont
  {Arkani-Hamed}}, \bibinfo {author} {\bibfnamefont {Y.-t.}\ \bibnamefont
  {Huang}}, \ and\ \bibinfo {author} {\bibfnamefont {D.}~\bibnamefont
  {O'Connell}},\ }\href {\doibase 10.1007/JHEP01(2020)046} {\bibfield
  {journal} {\bibinfo  {journal} {JHEP}\ }\textbf {\bibinfo {volume} {01}},\
  \bibinfo {pages} {046} (\bibinfo {year} {2020})},\ \Eprint
  {http://arxiv.org/abs/1906.10100} {arXiv:1906.10100 [hep-th]} \BibitemShut
  {NoStop}%
\bibitem [{\citenamefont {Bern}\ \emph {et~al.}(2021)\citenamefont {Bern},
  \citenamefont {Luna}, \citenamefont {Roiban}, \citenamefont {Shen},\ and\
  \citenamefont {Zeng}}]{Bern:2020buy}%
  \BibitemOpen
  \bibfield  {author} {\bibinfo {author} {\bibfnamefont {Z.}~\bibnamefont
  {Bern}}, \bibinfo {author} {\bibfnamefont {A.}~\bibnamefont {Luna}}, \bibinfo
  {author} {\bibfnamefont {R.}~\bibnamefont {Roiban}}, \bibinfo {author}
  {\bibfnamefont {C.-H.}\ \bibnamefont {Shen}}, \ and\ \bibinfo {author}
  {\bibfnamefont {M.}~\bibnamefont {Zeng}},\ }\href {\doibase
  10.1103/PhysRevD.104.065014} {\bibfield  {journal} {\bibinfo  {journal}
  {Phys. Rev. D}\ }\textbf {\bibinfo {volume} {104}},\ \bibinfo {pages}
  {065014} (\bibinfo {year} {2021})},\ \Eprint
  {http://arxiv.org/abs/2005.03071} {arXiv:2005.03071 [hep-th]} \BibitemShut
  {NoStop}%
\bibitem [{\citenamefont {Aoude}\ and\ \citenamefont
  {Ochirov}(2021)}]{Aoude:2021oqj}%
  \BibitemOpen
  \bibfield  {author} {\bibinfo {author} {\bibfnamefont {R.}~\bibnamefont
  {Aoude}}\ and\ \bibinfo {author} {\bibfnamefont {A.}~\bibnamefont
  {Ochirov}},\ }\href {\doibase 10.1007/JHEP10(2021)008} {\bibfield  {journal}
  {\bibinfo  {journal} {JHEP}\ }\textbf {\bibinfo {volume} {10}},\ \bibinfo
  {pages} {008} (\bibinfo {year} {2021})},\ \Eprint
  {http://arxiv.org/abs/2108.01649} {arXiv:2108.01649 [hep-th]} \BibitemShut
  {NoStop}%
\bibitem [{\citenamefont {Bergshoeff}\ \emph {et~al.}(2009)\citenamefont
  {Bergshoeff}, \citenamefont {Hohm},\ and\ \citenamefont
  {Townsend}}]{Bergshoeff:2009hq}%
  \BibitemOpen
  \bibfield  {author} {\bibinfo {author} {\bibfnamefont {E.~A.}\ \bibnamefont
  {Bergshoeff}}, \bibinfo {author} {\bibfnamefont {O.}~\bibnamefont {Hohm}}, \
  and\ \bibinfo {author} {\bibfnamefont {P.~K.}\ \bibnamefont {Townsend}},\
  }\href {\doibase 10.1103/PhysRevLett.102.201301} {\bibfield  {journal}
  {\bibinfo  {journal} {Phys. Rev. Lett.}\ }\textbf {\bibinfo {volume} {102}},\
  \bibinfo {pages} {201301} (\bibinfo {year} {2009})},\ \Eprint
  {http://arxiv.org/abs/0901.1766} {arXiv:0901.1766 [hep-th]} \BibitemShut
  {NoStop}%
\bibitem [{\citenamefont {de~Rham}\ \emph {et~al.}(2011)\citenamefont
  {de~Rham}, \citenamefont {Gabadadze},\ and\ \citenamefont
  {Tolley}}]{deRham:2010kj}%
  \BibitemOpen
  \bibfield  {author} {\bibinfo {author} {\bibfnamefont {C.}~\bibnamefont
  {de~Rham}}, \bibinfo {author} {\bibfnamefont {G.}~\bibnamefont {Gabadadze}},
  \ and\ \bibinfo {author} {\bibfnamefont {A.~J.}\ \bibnamefont {Tolley}},\
  }\href {\doibase 10.1103/PhysRevLett.106.231101} {\bibfield  {journal}
  {\bibinfo  {journal} {Phys. Rev. Lett.}\ }\textbf {\bibinfo {volume} {106}},\
  \bibinfo {pages} {231101} (\bibinfo {year} {2011})},\ \Eprint
  {http://arxiv.org/abs/1011.1232} {arXiv:1011.1232 [hep-th]} \BibitemShut
  {NoStop}%
\bibitem [{\citenamefont {Hassan}\ and\ \citenamefont
  {Rosen}(2012)}]{Hassan:2011zd}%
  \BibitemOpen
  \bibfield  {author} {\bibinfo {author} {\bibfnamefont {S.~F.}\ \bibnamefont
  {Hassan}}\ and\ \bibinfo {author} {\bibfnamefont {R.~A.}\ \bibnamefont
  {Rosen}},\ }\href {\doibase 10.1007/JHEP02(2012)126} {\bibfield  {journal}
  {\bibinfo  {journal} {JHEP}\ }\textbf {\bibinfo {volume} {02}},\ \bibinfo
  {pages} {126} (\bibinfo {year} {2012})},\ \Eprint
  {http://arxiv.org/abs/1109.3515} {arXiv:1109.3515 [hep-th]} \BibitemShut
  {NoStop}%
\bibitem [{\citenamefont {de~Rham}(2014)}]{deRham:2014zqa}%
  \BibitemOpen
  \bibfield  {author} {\bibinfo {author} {\bibfnamefont {C.}~\bibnamefont
  {de~Rham}},\ }\href {\doibase 10.12942/lrr-2014-7} {\bibfield  {journal}
  {\bibinfo  {journal} {Living Rev. Rel.}\ }\textbf {\bibinfo {volume} {17}},\
  \bibinfo {pages} {7} (\bibinfo {year} {2014})},\ \Eprint
  {http://arxiv.org/abs/1401.4173} {arXiv:1401.4173 [hep-th]} \BibitemShut
  {NoStop}%
\bibitem [{\citenamefont {Zinoviev}(2007)}]{Zinoviev:2006im}%
  \BibitemOpen
  \bibfield  {author} {\bibinfo {author} {\bibfnamefont {Y.~M.}\ \bibnamefont
  {Zinoviev}},\ }\href {\doibase 10.1016/j.nuclphysb.2007.02.005} {\bibfield
  {journal} {\bibinfo  {journal} {Nucl. Phys. B}\ }\textbf {\bibinfo {volume}
  {770}},\ \bibinfo {pages} {83} (\bibinfo {year} {2007})},\ \Eprint
  {http://arxiv.org/abs/hep-th/0609170} {arXiv:hep-th/0609170} \BibitemShut
  {NoStop}%
\bibitem [{\citenamefont {Zinoviev}(2009{\natexlab{a}})}]{Zinoviev:2008ck}%
  \BibitemOpen
  \bibfield  {author} {\bibinfo {author} {\bibfnamefont {Y.~M.}\ \bibnamefont
  {Zinoviev}},\ }\href {\doibase 10.1088/0264-9381/26/3/035022} {\bibfield
  {journal} {\bibinfo  {journal} {Class. Quant. Grav.}\ }\textbf {\bibinfo
  {volume} {26}},\ \bibinfo {pages} {035022} (\bibinfo {year}
  {2009}{\natexlab{a}})},\ \Eprint {http://arxiv.org/abs/0805.2226}
  {arXiv:0805.2226 [hep-th]} \BibitemShut {NoStop}%
\bibitem [{\citenamefont {Zinoviev}(2009{\natexlab{b}})}]{Zinoviev:2009hu}%
  \BibitemOpen
  \bibfield  {author} {\bibinfo {author} {\bibfnamefont {Y.~M.}\ \bibnamefont
  {Zinoviev}},\ }\href {\doibase 10.1016/j.nuclphysb.2009.04.027} {\bibfield
  {journal} {\bibinfo  {journal} {Nucl. Phys. B}\ }\textbf {\bibinfo {volume}
  {821}},\ \bibinfo {pages} {431} (\bibinfo {year} {2009}{\natexlab{b}})},\
  \Eprint {http://arxiv.org/abs/0901.3462} {arXiv:0901.3462 [hep-th]}
  \BibitemShut {NoStop}%
\bibitem [{\citenamefont {Zinoviev}(2011)}]{Zinoviev:2010av}%
  \BibitemOpen
  \bibfield  {author} {\bibinfo {author} {\bibfnamefont {Y.~M.}\ \bibnamefont
  {Zinoviev}},\ }\href {\doibase 10.1007/JHEP03(2011)082} {\bibfield  {journal}
  {\bibinfo  {journal} {JHEP}\ }\textbf {\bibinfo {volume} {03}},\ \bibinfo
  {pages} {082} (\bibinfo {year} {2011})},\ \Eprint
  {http://arxiv.org/abs/1012.2706} {arXiv:1012.2706 [hep-th]} \BibitemShut
  {NoStop}%
\bibitem [{\citenamefont {Zinoviev}(2012)}]{Zinoviev:2011fv}%
  \BibitemOpen
  \bibfield  {author} {\bibinfo {author} {\bibfnamefont {Y.~M.}\ \bibnamefont
  {Zinoviev}},\ }\href {\doibase 10.1088/0264-9381/29/1/015013} {\bibfield
  {journal} {\bibinfo  {journal} {Class. Quant. Grav.}\ }\textbf {\bibinfo
  {volume} {29}},\ \bibinfo {pages} {015013} (\bibinfo {year} {2012})},\
  \Eprint {http://arxiv.org/abs/1107.3222} {arXiv:1107.3222 [hep-th]}
  \BibitemShut {NoStop}%
\bibitem [{\citenamefont {Buchbinder}\ \emph {et~al.}(2013)\citenamefont
  {Buchbinder}, \citenamefont {Snegirev},\ and\ \citenamefont
  {Zinoviev}}]{Buchbinder:2012xa}%
  \BibitemOpen
  \bibfield  {author} {\bibinfo {author} {\bibfnamefont {I.~L.}\ \bibnamefont
  {Buchbinder}}, \bibinfo {author} {\bibfnamefont {T.~V.}\ \bibnamefont
  {Snegirev}}, \ and\ \bibinfo {author} {\bibfnamefont {Y.~M.}\ \bibnamefont
  {Zinoviev}},\ }\href {\doibase 10.1088/1751-8113/46/21/214015} {\bibfield
  {journal} {\bibinfo  {journal} {J. Phys. A}\ }\textbf {\bibinfo {volume}
  {46}},\ \bibinfo {pages} {214015} (\bibinfo {year} {2013})},\ \Eprint
  {http://arxiv.org/abs/1208.0183} {arXiv:1208.0183 [hep-th]} \BibitemShut
  {NoStop}%
\bibitem [{\citenamefont {Buchbinder}\ and\ \citenamefont
  {Krykhtin}(2005)}]{Buchbinder:2005ua}%
  \BibitemOpen
  \bibfield  {author} {\bibinfo {author} {\bibfnamefont {I.~L.}\ \bibnamefont
  {Buchbinder}}\ and\ \bibinfo {author} {\bibfnamefont {V.~A.}\ \bibnamefont
  {Krykhtin}},\ }\href {\doibase 10.1016/j.nuclphysb.2005.07.035} {\bibfield
  {journal} {\bibinfo  {journal} {Nucl. Phys. B}\ }\textbf {\bibinfo {volume}
  {727}},\ \bibinfo {pages} {537} (\bibinfo {year} {2005})},\ \Eprint
  {http://arxiv.org/abs/hep-th/0505092} {arXiv:hep-th/0505092} \BibitemShut
  {NoStop}%
\bibitem [{\citenamefont {Buchbinder}\ \emph {et~al.}(2007)\citenamefont
  {Buchbinder}, \citenamefont {Krykhtin},\ and\ \citenamefont
  {Takata}}]{Buchbinder:2007ix}%
  \BibitemOpen
  \bibfield  {author} {\bibinfo {author} {\bibfnamefont {I.~L.}\ \bibnamefont
  {Buchbinder}}, \bibinfo {author} {\bibfnamefont {V.~A.}\ \bibnamefont
  {Krykhtin}}, \ and\ \bibinfo {author} {\bibfnamefont {H.}~\bibnamefont
  {Takata}},\ }\href {\doibase 10.1016/j.physletb.2007.09.033} {\bibfield
  {journal} {\bibinfo  {journal} {Phys. Lett. B}\ }\textbf {\bibinfo {volume}
  {656}},\ \bibinfo {pages} {253} (\bibinfo {year} {2007})},\ \Eprint
  {http://arxiv.org/abs/0707.2181} {arXiv:0707.2181 [hep-th]} \BibitemShut
  {NoStop}%
\bibitem [{\citenamefont {Kaparulin}\ \emph {et~al.}(2013)\citenamefont
  {Kaparulin}, \citenamefont {Lyakhovich},\ and\ \citenamefont
  {Sharapov}}]{Kaparulin:2012px}%
  \BibitemOpen
  \bibfield  {author} {\bibinfo {author} {\bibfnamefont {D.~S.}\ \bibnamefont
  {Kaparulin}}, \bibinfo {author} {\bibfnamefont {S.~L.}\ \bibnamefont
  {Lyakhovich}}, \ and\ \bibinfo {author} {\bibfnamefont {A.~A.}\ \bibnamefont
  {Sharapov}},\ }\href {\doibase 10.1007/JHEP01(2013)097} {\bibfield  {journal}
  {\bibinfo  {journal} {JHEP}\ }\textbf {\bibinfo {volume} {01}},\ \bibinfo
  {pages} {097} (\bibinfo {year} {2013})},\ \Eprint
  {http://arxiv.org/abs/1210.6821} {arXiv:1210.6821 [hep-th]} \BibitemShut
  {NoStop}%
\bibitem [{\citenamefont {Kazinski}\ \emph {et~al.}(2005)\citenamefont
  {Kazinski}, \citenamefont {Lyakhovich},\ and\ \citenamefont
  {Sharapov}}]{Kazinski:2005eb}%
  \BibitemOpen
  \bibfield  {author} {\bibinfo {author} {\bibfnamefont {P.~O.}\ \bibnamefont
  {Kazinski}}, \bibinfo {author} {\bibfnamefont {S.~L.}\ \bibnamefont
  {Lyakhovich}}, \ and\ \bibinfo {author} {\bibfnamefont {A.~A.}\ \bibnamefont
  {Sharapov}},\ }\href {\doibase 10.1088/1126-6708/2005/07/076} {\bibfield
  {journal} {\bibinfo  {journal} {JHEP}\ }\textbf {\bibinfo {volume} {07}},\
  \bibinfo {pages} {076} (\bibinfo {year} {2005})},\ \Eprint
  {http://arxiv.org/abs/hep-th/0506093} {arXiv:hep-th/0506093} \BibitemShut
  {NoStop}%
\bibitem [{\citenamefont {Metsaev}(2006)}]{Metsaev:2005ar}%
  \BibitemOpen
  \bibfield  {author} {\bibinfo {author} {\bibfnamefont {R.~R.}\ \bibnamefont
  {Metsaev}},\ }\href {\doibase 10.1016/j.nuclphysb.2006.10.002} {\bibfield
  {journal} {\bibinfo  {journal} {Nucl. Phys. B}\ }\textbf {\bibinfo {volume}
  {759}},\ \bibinfo {pages} {147} (\bibinfo {year} {2006})},\ \Eprint
  {http://arxiv.org/abs/hep-th/0512342} {arXiv:hep-th/0512342} \BibitemShut
  {NoStop}%
\bibitem [{\citenamefont {Metsaev}(2012)}]{Metsaev:2007rn}%
  \BibitemOpen
  \bibfield  {author} {\bibinfo {author} {\bibfnamefont {R.~R.}\ \bibnamefont
  {Metsaev}},\ }\href {\doibase 10.1016/j.nuclphysb.2012.01.022} {\bibfield
  {journal} {\bibinfo  {journal} {Nucl. Phys. B}\ }\textbf {\bibinfo {volume}
  {859}},\ \bibinfo {pages} {13} (\bibinfo {year} {2012})},\ \Eprint
  {http://arxiv.org/abs/0712.3526} {arXiv:0712.3526 [hep-th]} \BibitemShut
  {NoStop}%
\bibitem [{\citenamefont {Metsaev}(2022)}]{Metsaev:2022yvb}%
  \BibitemOpen
  \bibfield  {author} {\bibinfo {author} {\bibfnamefont {R.~R.}\ \bibnamefont
  {Metsaev}},\ }\href@noop {} {\  (\bibinfo {year} {2022})},\ \Eprint
  {http://arxiv.org/abs/2206.13268} {arXiv:2206.13268 [hep-th]} \BibitemShut
  {NoStop}%
\bibitem [{\citenamefont {Albonico}\ \emph {et~al.}(2022)\citenamefont
  {Albonico}, \citenamefont {Geyer},\ and\ \citenamefont
  {Mason}}]{Albonico:2022pmd}%
  \BibitemOpen
  \bibfield  {author} {\bibinfo {author} {\bibfnamefont {G.}~\bibnamefont
  {Albonico}}, \bibinfo {author} {\bibfnamefont {Y.}~\bibnamefont {Geyer}}, \
  and\ \bibinfo {author} {\bibfnamefont {L.}~\bibnamefont {Mason}},\ }\href
  {\doibase 10.3842/SIGMA.2022.045} {\bibfield  {journal} {\bibinfo  {journal}
  {SIGMA}\ }\textbf {\bibinfo {volume} {18}},\ \bibinfo {pages} {045} (\bibinfo
  {year} {2022})},\ \Eprint {http://arxiv.org/abs/2203.08087} {arXiv:2203.08087
  [hep-th]} \BibitemShut {NoStop}%
\bibitem [{\citenamefont {Falkowski}\ and\ \citenamefont
  {Machado}(2021)}]{Falkowski:2020aso}%
  \BibitemOpen
  \bibfield  {author} {\bibinfo {author} {\bibfnamefont {A.}~\bibnamefont
  {Falkowski}}\ and\ \bibinfo {author} {\bibfnamefont {C.~S.}\ \bibnamefont
  {Machado}},\ }\href {\doibase 10.1007/JHEP05(2021)238} {\bibfield  {journal}
  {\bibinfo  {journal} {JHEP}\ }\textbf {\bibinfo {volume} {05}},\ \bibinfo
  {pages} {238} (\bibinfo {year} {2021})},\ \Eprint
  {http://arxiv.org/abs/2005.08981} {arXiv:2005.08981 [hep-th]} \BibitemShut
  {NoStop}%
\bibitem [{\citenamefont {Chiodaroli}\ \emph {et~al.}(2022)\citenamefont
  {Chiodaroli}, \citenamefont {Johansson},\ and\ \citenamefont
  {Pichini}}]{Chiodaroli:2021eug}%
  \BibitemOpen
  \bibfield  {author} {\bibinfo {author} {\bibfnamefont {M.}~\bibnamefont
  {Chiodaroli}}, \bibinfo {author} {\bibfnamefont {H.}~\bibnamefont
  {Johansson}}, \ and\ \bibinfo {author} {\bibfnamefont {P.}~\bibnamefont
  {Pichini}},\ }\href {\doibase 10.1007/JHEP02(2022)156} {\bibfield  {journal}
  {\bibinfo  {journal} {JHEP}\ }\textbf {\bibinfo {volume} {02}},\ \bibinfo
  {pages} {156} (\bibinfo {year} {2022})},\ \Eprint
  {http://arxiv.org/abs/2107.14779} {arXiv:2107.14779 [hep-th]} \BibitemShut
  {NoStop}%
\bibitem [{\citenamefont {Penrose}(1965)}]{Penrose:1965am}%
  \BibitemOpen
  \bibfield  {author} {\bibinfo {author} {\bibfnamefont {R.}~\bibnamefont
  {Penrose}},\ }\href {\doibase 10.1098/rspa.1965.0058} {\bibfield  {journal}
  {\bibinfo  {journal} {Proc. Roy. Soc. Lond.}\ }\textbf {\bibinfo {volume}
  {A284}},\ \bibinfo {pages} {159} (\bibinfo {year} {1965})}\BibitemShut
  {NoStop}%
\bibitem [{\citenamefont {Hughston}\ \emph {et~al.}(1979)\citenamefont
  {Hughston}, \citenamefont {Ward}, \citenamefont {Eastwood}, \citenamefont
  {Ginsberg}, \citenamefont {Hodges}, \citenamefont {Huggett}, \citenamefont
  {Hurd}, \citenamefont {Jozsa}, \citenamefont {Penrose}, \citenamefont
  {Popovich} \emph {et~al.}}]{Hughston:1979tq}%
  \BibitemOpen
  \bibinfo {editor} {\bibfnamefont {L.~P.}\ \bibnamefont {Hughston}}, \bibinfo
  {editor} {\bibfnamefont {R.~S.}\ \bibnamefont {Ward}}, \bibinfo {editor}
  {\bibfnamefont {M.~G.}\ \bibnamefont {Eastwood}}, \bibinfo {editor}
  {\bibfnamefont {M.~L.}\ \bibnamefont {Ginsberg}}, \bibinfo {editor}
  {\bibfnamefont {A.~P.}\ \bibnamefont {Hodges}}, \bibinfo {editor}
  {\bibfnamefont {S.~A.}\ \bibnamefont {Huggett}}, \bibinfo {editor}
  {\bibfnamefont {T.~R.}\ \bibnamefont {Hurd}}, \bibinfo {editor}
  {\bibfnamefont {R.~O.}\ \bibnamefont {Jozsa}}, \bibinfo {editor}
  {\bibfnamefont {R.}~\bibnamefont {Penrose}}, \bibinfo {editor} {\bibfnamefont
  {A.}~\bibnamefont {Popovich}},  \emph {et~al.},\ eds.,\ \href@noop {} {\emph
  {\bibinfo {title} {{Advances in twistor theory}}}}\ (\bibinfo {year}
  {1979})\BibitemShut {NoStop}%
\bibitem [{\citenamefont {Eastwood}\ \emph {et~al.}(1981)\citenamefont
  {Eastwood}, \citenamefont {Penrose},\ and\ \citenamefont
  {Wells}}]{Eastwood:1981jy}%
  \BibitemOpen
  \bibfield  {author} {\bibinfo {author} {\bibfnamefont {M.~G.}\ \bibnamefont
  {Eastwood}}, \bibinfo {author} {\bibfnamefont {R.}~\bibnamefont {Penrose}}, \
  and\ \bibinfo {author} {\bibfnamefont {R.~O.}\ \bibnamefont {Wells}},\ }\href
  {\doibase 10.1007/BF01942327} {\bibfield  {journal} {\bibinfo  {journal}
  {Commun. Math. Phys.}\ }\textbf {\bibinfo {volume} {78}},\ \bibinfo {pages}
  {305} (\bibinfo {year} {1981})}\BibitemShut {NoStop}%
\bibitem [{\citenamefont {Woodhouse}(1985)}]{Woodhouse:1985id}%
  \BibitemOpen
  \bibfield  {author} {\bibinfo {author} {\bibfnamefont {N.~M.~J.}\
  \bibnamefont {Woodhouse}},\ }\href {\doibase 10.1088/0264-9381/2/3/006}
  {\bibfield  {journal} {\bibinfo  {journal} {Class. Quant. Grav.}\ }\textbf
  {\bibinfo {volume} {2}},\ \bibinfo {pages} {257} (\bibinfo {year}
  {1985})}\BibitemShut {NoStop}%
\bibitem [{\citenamefont {Huggett}\ and\ \citenamefont
  {Tod}(1986)}]{Huggett:1986fs}%
  \BibitemOpen
  \bibfield  {author} {\bibinfo {author} {\bibfnamefont {S.~A.}\ \bibnamefont
  {Huggett}}\ and\ \bibinfo {author} {\bibfnamefont {K.~P.}\ \bibnamefont
  {Tod}},\ }\href {\doibase 10.1017/CBO9780511624018} {\emph {\bibinfo {title}
  {{An Introduction to Twistor Theory}}}},\ London Mathematical Society Student
  Texts\ (\bibinfo  {publisher} {Cambridge University Press},\ \bibinfo {year}
  {1986})\BibitemShut {NoStop}%
\bibitem [{\citenamefont {Skvortsov}\ and\ \citenamefont
  {Van~Dongen}(2022)}]{Skvortsov:2022syz}%
  \BibitemOpen
  \bibfield  {author} {\bibinfo {author} {\bibfnamefont {E.}~\bibnamefont
  {Skvortsov}}\ and\ \bibinfo {author} {\bibfnamefont {R.}~\bibnamefont
  {Van~Dongen}},\ }\href@noop {} {\  (\bibinfo {year} {2022})},\ \Eprint
  {http://arxiv.org/abs/2204.10285} {arXiv:2204.10285 [hep-th]} \BibitemShut
  {NoStop}%
\bibitem [{\citenamefont {Chalmers}\ and\ \citenamefont
  {Siegel}(1999)}]{Chalmers:1997ui}%
  \BibitemOpen
  \bibfield  {author} {\bibinfo {author} {\bibfnamefont {G.}~\bibnamefont
  {Chalmers}}\ and\ \bibinfo {author} {\bibfnamefont {W.}~\bibnamefont
  {Siegel}},\ }\href {\doibase 10.1103/PhysRevD.59.045012} {\bibfield
  {journal} {\bibinfo  {journal} {Phys. Rev.}\ }\textbf {\bibinfo {volume}
  {D59}},\ \bibinfo {pages} {045012} (\bibinfo {year} {1999})},\ \Eprint
  {http://arxiv.org/abs/hep-ph/9708251} {arXiv:hep-ph/9708251 [hep-ph]}
  \BibitemShut {NoStop}%
\bibitem [{\citenamefont {Chalmers}\ and\ \citenamefont
  {Siegel}(2001)}]{Chalmers:2001cy}%
  \BibitemOpen
  \bibfield  {author} {\bibinfo {author} {\bibfnamefont {G.}~\bibnamefont
  {Chalmers}}\ and\ \bibinfo {author} {\bibfnamefont {W.}~\bibnamefont
  {Siegel}},\ }\href {\doibase 10.1103/PhysRevD.63.125027} {\bibfield
  {journal} {\bibinfo  {journal} {Phys. Rev.}\ }\textbf {\bibinfo {volume}
  {D63}},\ \bibinfo {pages} {125027} (\bibinfo {year} {2001})},\ \Eprint
  {http://arxiv.org/abs/hep-th/0101025} {arXiv:hep-th/0101025 [hep-th]}
  \BibitemShut {NoStop}%
\bibitem [{\citenamefont {Fradkin}\ and\ \citenamefont
  {Tseytlin}(1985)}]{Fradkin:1984ai}%
  \BibitemOpen
  \bibfield  {author} {\bibinfo {author} {\bibfnamefont {E.~S.}\ \bibnamefont
  {Fradkin}}\ and\ \bibinfo {author} {\bibfnamefont {A.~A.}\ \bibnamefont
  {Tseytlin}},\ }\href {\doibase 10.1016/0003-4916(85)90225-8} {\bibfield
  {journal} {\bibinfo  {journal} {Annals Phys.}\ }\textbf {\bibinfo {volume}
  {162}},\ \bibinfo {pages} {31} (\bibinfo {year} {1985})}\BibitemShut
  {NoStop}%
\bibitem [{\citenamefont {Krasnov}(2017)}]{Krasnov:2016emc}%
  \BibitemOpen
  \bibfield  {author} {\bibinfo {author} {\bibfnamefont {K.}~\bibnamefont
  {Krasnov}},\ }\href {\doibase 10.1088/1361-6382/aa65e5} {\bibfield  {journal}
  {\bibinfo  {journal} {Class. Quant. Grav.}\ }\textbf {\bibinfo {volume}
  {34}},\ \bibinfo {pages} {095001} (\bibinfo {year} {2017})},\ \Eprint
  {http://arxiv.org/abs/1610.01457} {arXiv:1610.01457 [hep-th]} \BibitemShut
  {NoStop}%
\bibitem [{\citenamefont {Levi}\ and\ \citenamefont
  {Steinhoff}(2015)}]{Levi:2015msa}%
  \BibitemOpen
  \bibfield  {author} {\bibinfo {author} {\bibfnamefont {M.}~\bibnamefont
  {Levi}}\ and\ \bibinfo {author} {\bibfnamefont {J.}~\bibnamefont
  {Steinhoff}},\ }\href {\doibase 10.1007/JHEP09(2015)219} {\bibfield
  {journal} {\bibinfo  {journal} {JHEP}\ }\textbf {\bibinfo {volume} {09}},\
  \bibinfo {pages} {219} (\bibinfo {year} {2015})},\ \Eprint
  {http://arxiv.org/abs/1501.04956} {arXiv:1501.04956 [gr-qc]} \BibitemShut
  {NoStop}%
\bibitem [{\citenamefont {Porto}(2016)}]{Porto:2016pyg}%
  \BibitemOpen
  \bibfield  {author} {\bibinfo {author} {\bibfnamefont {R.~A.}\ \bibnamefont
  {Porto}},\ }\href {\doibase 10.1016/j.physrep.2016.04.003} {\bibfield
  {journal} {\bibinfo  {journal} {Phys. Rept.}\ }\textbf {\bibinfo {volume}
  {633}},\ \bibinfo {pages} {1} (\bibinfo {year} {2016})},\ \Eprint
  {http://arxiv.org/abs/1601.04914} {arXiv:1601.04914 [hep-th]} \BibitemShut
  {NoStop}%
\bibitem [{\citenamefont {Vines}(2018)}]{Vines:2017hyw}%
  \BibitemOpen
  \bibfield  {author} {\bibinfo {author} {\bibfnamefont {J.}~\bibnamefont
  {Vines}},\ }\href {\doibase 10.1088/1361-6382/aaa3a8} {\bibfield  {journal}
  {\bibinfo  {journal} {Class. Quant. Grav.}\ }\textbf {\bibinfo {volume}
  {35}},\ \bibinfo {pages} {084002} (\bibinfo {year} {2018})},\ \Eprint
  {http://arxiv.org/abs/1709.06016} {arXiv:1709.06016 [gr-qc]} \BibitemShut
  {NoStop}%
\bibitem [{\citenamefont {Guevara}\ \emph {et~al.}(2021)\citenamefont
  {Guevara}, \citenamefont {Maybee}, \citenamefont {Ochirov}, \citenamefont
  {O'Connell},\ and\ \citenamefont {Vines}}]{Guevara:2020xjx}%
  \BibitemOpen
  \bibfield  {author} {\bibinfo {author} {\bibfnamefont {A.}~\bibnamefont
  {Guevara}}, \bibinfo {author} {\bibfnamefont {B.}~\bibnamefont {Maybee}},
  \bibinfo {author} {\bibfnamefont {A.}~\bibnamefont {Ochirov}}, \bibinfo
  {author} {\bibfnamefont {D.}~\bibnamefont {O'Connell}}, \ and\ \bibinfo
  {author} {\bibfnamefont {J.}~\bibnamefont {Vines}},\ }\href {\doibase
  10.1007/JHEP03(2021)201} {\bibfield  {journal} {\bibinfo  {journal} {JHEP}\
  }\textbf {\bibinfo {volume} {03}},\ \bibinfo {pages} {201} (\bibinfo {year}
  {2021})},\ \Eprint {http://arxiv.org/abs/2012.11570} {arXiv:2012.11570
  [hep-th]} \BibitemShut {NoStop}%
\bibitem [{\citenamefont {Holstein}\ and\ \citenamefont
  {Ross}(2008{\natexlab{a}})}]{Holstein:2008sw}%
  \BibitemOpen
  \bibfield  {author} {\bibinfo {author} {\bibfnamefont {B.~R.}\ \bibnamefont
  {Holstein}}\ and\ \bibinfo {author} {\bibfnamefont {A.}~\bibnamefont
  {Ross}},\ }\href@noop {} {\  (\bibinfo {year} {2008}{\natexlab{a}})},\
  \Eprint {http://arxiv.org/abs/0802.0715} {arXiv:0802.0715 [hep-ph]}
  \BibitemShut {NoStop}%
\bibitem [{\citenamefont {Holstein}\ and\ \citenamefont
  {Ross}(2008{\natexlab{b}})}]{Holstein:2008sx}%
  \BibitemOpen
  \bibfield  {author} {\bibinfo {author} {\bibfnamefont {B.~R.}\ \bibnamefont
  {Holstein}}\ and\ \bibinfo {author} {\bibfnamefont {A.}~\bibnamefont
  {Ross}},\ }\href@noop {} {\  (\bibinfo {year} {2008}{\natexlab{b}})},\
  \Eprint {http://arxiv.org/abs/0802.0716} {arXiv:0802.0716 [hep-ph]}
  \BibitemShut {NoStop}%
\bibitem [{\citenamefont {Vaidya}(2015)}]{Vaidya:2014kza}%
  \BibitemOpen
  \bibfield  {author} {\bibinfo {author} {\bibfnamefont {V.}~\bibnamefont
  {Vaidya}},\ }\href {\doibase 10.1103/PhysRevD.91.024017} {\bibfield
  {journal} {\bibinfo  {journal} {Phys. Rev.}\ }\textbf {\bibinfo {volume}
  {D91}},\ \bibinfo {pages} {024017} (\bibinfo {year} {2015})},\ \Eprint
  {http://arxiv.org/abs/1410.5348} {arXiv:1410.5348 [hep-th]} \BibitemShut
  {NoStop}%
\bibitem [{\citenamefont {Cangemi}\ and\ \citenamefont
  {Pichini}(2022)}]{Cangemi:2022abk}%
  \BibitemOpen
  \bibfield  {author} {\bibinfo {author} {\bibfnamefont {L.}~\bibnamefont
  {Cangemi}}\ and\ \bibinfo {author} {\bibfnamefont {P.}~\bibnamefont
  {Pichini}},\ }\href@noop {} {\  (\bibinfo {year} {2022})},\ \Eprint
  {http://arxiv.org/abs/2207.03947} {arXiv:2207.03947 [hep-th]} \BibitemShut
  {NoStop}%
\end{thebibliography}%
\end{document}